\documentclass[secnumarabic,amssymb, nofootinbib, notitlepage, 11pt, aps, prd]{revtex4-1}
\usepackage{graphicx,xcolor}  % needed for figures
\usepackage{dcolumn}   % needed for some tables
\usepackage{bm}        % for math
\usepackage{amssymb}   % for math
\usepackage{bbding}
\usepackage{amsfonts}
\usepackage{bbm}
\usepackage{dsfont}
\usepackage{hyperref}
\usepackage[toc,page,title,titletoc,header]{appendix}
\usepackage{mathrsfs}
\usepackage{amssymb,graphicx,units,yfonts}
\usepackage{amsmath}
\usepackage{txfonts}
\usepackage{setspace}
\usepackage{framed}

\begin{document}

\title{Lessons from $f(R,R_c^2,R_m^2,\mathscr{L}_m)$ gravity:
Smooth Gauss-Bonnet limit, energy-momentum conservation, and nonminimal coupling}%

%\author[1]{David W. Tian\thanks{Email address: wtian@mun.ca}}
%\author[2]{Ivan Booth\thanks{Email address: ibooth@mun.ca}}
%\affil[1]{ Faculty of Science,  Memorial University, St. John's, NL, Canada, A1C 5S7}
%\affil[2]{ Department of Mathematics and Statistics, Memorial University, St. John's,  NL, Canada, A1C 5S7}
\author{David W. Tian}%
\email[]{Email address: wtian@mun.ca}
\affiliation{Faculty of Science,  Memorial University, St. John's, Newfoundland, Canada, A1C 5S7}
\author{Ivan Booth}%
\email[]{Email address: ibooth@mun.ca}
\affiliation{Department of Mathematics and Statistics, Memorial University, St. John's,  Newfoundland, Canada, A1C 5S7}
%\date{}%

\begin{abstract}
This paper studies a  generic fourth-order theory of gravity with Lagrangian density $f(R,R_c^2,R_m^2, \mathscr{L}_m)$, where $R_c^2$ and $R_m^2$ respectively denote the square of the Ricci and Riemann tensors. By considering explicit $R^2$
dependence and imposing the ``coherence condition'' $f_{R^2}\!=\!f_{R_m^2}\!=\! -f_{R_c^2}/4$, the field equations of $f(R,R^2,R_c^2,R_m^2, \mathscr{L}_m)$ gravity can be smoothly reduced to that of $f(R,\mathcal{G},\mathscr{L}_m)$ generalized Gauss-Bonnet gravity with $\mathcal{G}$ denoting the Gauss-Bonnet invariant. We use Noether's conservation law to study the $f(\mathcal{R}_1,\mathcal{R}_2\ldots,\mathcal{R}_n,\mathscr{L}_m)$ model with nonminimal coupling between $\mathscr{L}_m$ and Riemannian invariants $ \mathcal{R}_i$, and conjecture that the gradient of nonminimal gravitational coupling strength $\nabla^\mu f_{\!\mathscr{L}_m}$ is the only source for energy-momentum nonconservation. This conjecture is applied to the $f(R,R_c^2,R_m^2, \mathscr{L}_m)$ model, and the equations of continuity and nongeodesic motion of different matter contents are investigated. Finally, the field equation for Lagrangians including the traceless-Ricci square and traceless-Riemann (Weyl) square invariants is derived, the $f(R,R_c^2,R_m^2,
\mathscr{L}_m)$ model is compared with the $f(R,R_c^2,R_m^2,T)+2\kappa \mathscr{L}_m$ model, and consequences of nonminimal coupling for black hole and wormhole physics are considered.\\

\end{abstract}

\maketitle
%\tableofcontents

%%%%%%%%%%%%%%%%%%%%%%%%%%%%%%%%%%%%%%%%%%%%%%%%%%%%%%%%%%%%%%%%%%%%%%%%%%%%%
%%%%%%%%%%%%%%%%%%%%%%%%%%%%%%%%%%%%%%%%%%%%%%%%%%%%%%%%%%%%%%%%%%%%%%%%%%%%%

%\vspace{-0.75cm}
\section{Introduction}\label{Section 1}

There are two main proposals to explain the  accelerated expansion of the Universe\cite{Dark Energy 1}.
The first assumes the existence of negative-pressure dark energy as a dominant component of the cosmos\cite{Dark Energy 2}\cite{Dark Energy 3}.
%with a (somewhat exotic) large negative pressure
The second approach
seeks viable modifications of both general relativity (GR)
and its alternatives\cite{Dark Energy 4}\cite{f(R) gravity Review by Sotiriou and Faraoni}.% \\
%Also, effort can even goes further in the meantime,
%trying to unify dark energy with dark matter\cite{Dark Energy Dark matter together}, and the late-time acceleration with
%initial inflation\cite{GaussBonnet Review}.% \\

Focusing on modifications of GR, the original Lagrangian density can be modified in two ways: (1) extending its dependence on the curvature invariants, and (2) considering nonminimal curvature-matter coupling. The simplest curvature-invariant modification is
$f(R) + 2\kappa \mathscr{L}_m$ gravity\cite{f(R) gravity Review by Sotiriou and Faraoni}\cite{GaussBonnet Review} ($\kappa = 8\pi G/c^4 \equiv  8\pi G$ and $c = 1$ hereafter), where the isolated Ricci scalar $R$ in the Hilbert-Einstein action is replaced by a generic function of $R$. In this case standard energy-momentum conservation $\nabla^\mu T_{\mu\nu}=0$ continues to hold.
Further extensions have introduced dependence on such things as the Gauss-Bonnet invariant $\mathcal{G}$\cite{Dark Energy 4}\cite{GaussBonnet first model R/2k+f(G)} and squares of Ricci and Riemann tensors  $\{R_c^2,R_m^2\}$\cite{Carroll R+ f(R Rc2 Rm2)+2kLm}, leading to models with Lagrangian densities like $R + f(\mathcal{G}) + 2\kappa \mathscr{L}_m$, $f(R,\mathcal{G}) + 2\kappa \mathscr{L}_m$ and $R + f(R,R_c^2,R_m^2) + 2\kappa \mathscr{L}_m$. In all these models, the spacetime geometry remains minimally coupled to the matter Lagrangian density $\mathscr{L}_m$. %\\

On the other hand, following the spirit of  nonminimal $f(R) \mathscr{L}_d$ coupling in  scalar-field dark-energy models\cite{Dark energy nonminimal couplings},  for modified theories of gravity an extra term $\lambda \tilde{f}(R) \mathscr{L}_m$  was respectively added to the standard actions of GR and $f(R)\!+\!2\kappa \mathscr{L}_m$ gravity in \cite{Generalized Bianchi and Conservation 2} and \cite{f(RLm) Extra Force}, which represents nonminimal curvature-matter coupling between $R$ and  $\mathscr{L}_m$. These ideas soon attracted a lot of attention in other modifications of GR after the work in \cite{f(RLm) Extra Force}, and nonminimal coupling was introduced to other gravity models such as generalized Gauss-Bonnet gravity\cite{GaussBonnet Review}\cite{Gauss-Bonnet with Nonminimal Coupling f(G)Lm} with terms like $\lambda f(\mathcal{G}) \mathscr{L}_m$.
From these initial models, some general consequences of nonminimal coupling were revealed. Most significantly, $\mathscr{L}_m$ enters the gravitational field
equation directly, nonminimal coupling violates the equivalence principle, and in general, energy-momentum conservation is violated with nontrivial
energy-momentum-curvature transformation. In \cite{f(R Lm)}, $f(R,\mathscr{L}_m)$ theory as the most generic extension of GR within the dependence of $\{R,\mathscr{L}_m\}$
was developed, while another type of nonminimal coupling, the $f(R,T)\!+\!2\kappa  \mathscr{L}_m$ model, was considered in \cite{f(R T)}.%\\

In this paper, we consider modifications to GR from both invariant-dependence and  nonminimal-coupling aspects,
and introduce a  new model of generic fourth-order gravity with Lagrangian density  $f(R,R_c^2,R_m^2, \mathscr{L}_m)$.
This can be regarded as a generalization of the $f(R,\mathscr{L}_m)$ model\cite{f(R Lm)} by adding $R_c^2$ and $R_m^2$dependence,
and an extension of the $f(R,R_c^2,R_m^2)\!+\!2\kappa\mathscr{L}_m$ model\cite{Carroll R+ f(R Rc2 Rm2)+2kLm} by
allowing nonminimal curvature-matter coupling. Among the fourteen  independent algebraic invariants which can be constructed from the Riemann tensor and metric tensor\cite{Riemann invariants}\cite{Riemann invariants II}, besides  $R$ we focus on Ricci square $R_c^2$
and Riemann square (Kretschmann scalar) $R_m^2$, not only because they are the two simplest square invariants (as opposed to cubic
and quartic invariants\cite{Riemann invariants II}), but also because they provide a bridge to generalized Gauss-Bonnet theories of gravity\cite{GaussBonnet Review} and quadratic gravity\cite{Gauss-Bonnet Simplest and Quadratic}\cite{Quardratic gravity second paper}.
By studying this model, we hope to get further insights into the effects of nonminimal coupling and dependence on extra curvature invariants.% \\

This paper is organized as follows. First of all, the field equations for $\mathscr{L}\!=\!f(R,R_c^2,R_m^2, \mathscr{L}_m)$ gravity are derived and nonminimal couplings with $\mathscr{L}_m$ and $T$  are compared in Sec.~\ref{Section 2}. In Sec.~\ref{Section 3}, we consider  an explicit dependence on $R^2$, and introduce the  condition $f_{R^2}\!=\!f_{R_m^2}\!=\!  -f_{R_c^2}/4 $ to smoothly transform  $f(R,R^2,R_c^2,R_m^2, \mathscr{L}_m)$ gravity to the generalized
Gauss-Bonnet gravity $\mathscr{L}\!=\!f(R, \mathcal{G}, \mathscr{L}_m)$; employing $\mathcal{G}$, quadratic gravity is revisited and traceless models like
$\mathscr{L}\!=\!f(R,R_S^2, \mathcal{C}^2, \mathscr{L}_m)$ are discussed. In Sec.~\ref{Section 4}, we commit ourselves to understanding the energy-momentum
divergence problem associated with $f(R,R_c^2,R_m^2, \mathscr{L}_m)$ gravity and most generic  $\mathscr{L}\!=\!f(\mathcal{R}_1,\mathcal{R}_2\ldots,\mathcal{R}_n,\mathscr{L}_m)$
gravity with nonminimal coupling, as an application of which, the equations of continuity and nongeodesic motion are derived in Sec.~\ref{Section 5}. Finally,
in Sec.~\ref{Section 6},  two implications of nonminimal coupling for black hole physics and  wormholes are discussed.
%Specifically we consider the implications of generalized energy conditions for wormholes and isolated horizons.
In the Appendix  generalized energy conditions of $f(R,\mathcal{R}_1,\mathcal{R}_2\ldots,\mathcal{R}_n,\mathscr{L}_m)$ and $f(R,R_c^2,R_m^2, \mathscr{L}_m)$ gravity are considered. Throughout this paper, we adopt the sign convention $R^{\alpha}_{\;\;\,\beta\gamma\delta}=\partial_\gamma \Gamma^\alpha_{\delta\beta}-\partial_\delta \Gamma^\alpha_{\gamma\beta}\cdots$ with the metric signature $(-,+++)$, and follow the straightforward metric approach rather than first-order Einstein-Palatini.

%%%%%%%%%%%%%%%%%%%%%%%%%%%%%%%%%%%%%%%%%%%%%%%%%%%%%%%%%%%%%%%%%%%%%%%%%%%%%
%%%%%%%%%%%%%%%%%%%%%%%%%%%%%%%%%%%%%%%%%%%%%%%%%%%%%%%%%%%%%%%%%%%%%%%%%%%%%

%\vspace{-0.26cm}
\section{Field equation and its properties}\label{Section 2}
%\vspace{-0.1cm}
\subsection{Action and field equations}
The action we propose for a generic fourth-order theory of gravity with possibly nonminimal curvature-matter
coupling\footnote{The terms \emph{geometry-matter} coupling and \emph{curvature-matter} coupling
are both used in this paper. They are not identical: the former can be either nonminimal or minimal,
while the latter by its name is always nonminimal since a curvature invariant contains at least second-order derivative of the metric tensor.
Here nonminimal coupling happens between algebraic or differential Riemannian scalar invariants and $\mathscr{L}_m$, so we will mainly use
curvature-matter coupling.} is
\begin{equation}\label{Action}
\mathcal{S}\,=\,\int d^4x \sqrt{-g}\,f(R\,,R_c^2\,,R_m^2\,, \mathscr{L}_m)\;,
\end{equation}
where $R_c^2$ and $R_m^2$ denote the square of Ricci and Riemann curvature tensor, respectively,
\begin{equation}
R_c^2\,\coloneqq\,R_{\alpha\beta}R^{\alpha\beta}\quad,\quad R_m^2\,\coloneqq\,R_{\alpha\mu\beta\nu}\,R^{\alpha\mu\beta\nu}\;.
\end{equation}
Varying the action Eq.(\ref{Action}) with respect to the inverse metric $g^{\mu\nu}$, we  get
\begin{equation}\label{ActionVariation 1}
\delta \mathcal{S}\,=\int d^4x \!\sqrt{-g}\,\left\{-\frac{1}{2}f\,g_{\mu\nu}\!\cdot \delta g^{\mu\nu}+f_R\!\cdot\!\delta R +f_{R_c^2}\!\cdot
\delta R_c^2 +f_{R_m^2}\!\cdot \delta R_m^2+ f_{\!\mathscr{L}_m}\!\cdot\delta\mathscr{L}_m \right\}\;,\\
\end{equation}
where $f_R \coloneqq \partial f/ \partial R$ , $f_{R_c^2}  \coloneqq  \partial f/\partial  R_c^2$ , $f_{R_m^2} \coloneqq  \partial f/\partial R_m^2$ , and
$f_{\!\mathscr{L}_m} \coloneqq \partial f/ \partial \mathscr{L}_m$ . $\delta R_c^2$ and $\delta R_m^2$ can be reduced into  variations of Riemann tensor,
\begin{align}\label{Variation Step 2}
\delta R_c^2\,=\,\delta \,\bigg[R_{\alpha\beta}  \!\cdot\! \big(g^{\alpha\rho}g^{\beta\sigma}R_{\rho\sigma}\big) \bigg]\,=\,2R_\mu^{\;\;\,\alpha}R_{\alpha\nu} \!\cdot \delta g^{\mu\nu} + 2R^{\mu\nu} \!\cdot \delta R^\alpha_{\;\;\,\mu\alpha\nu}\;,
\end{align}
\begin{align}\label{Variation Step 3}
\delta R_m^2\,=\,\delta \,\bigg[R_{\alpha\beta\gamma\epsilon} \!\cdot\! \big(g^{\alpha\rho}g^{\beta\sigma}g^{\gamma\zeta}g^{\epsilon\eta}R_{\rho\sigma\zeta\eta}\big)   \bigg]
\,=\,4R_{\mu\alpha\beta\gamma}R_{\nu}^{\;\;\,\alpha\beta\gamma} \!\cdot \delta g^{\mu\nu}
+2R^{\alpha\beta\gamma\epsilon} \!\cdot   \big(R^{\rho}_{\;\;\beta\gamma\epsilon} \,\delta g_{\alpha\rho} +g_{\alpha\rho} \,\delta  R^{\rho}_{\;\;\beta\gamma\epsilon} \big)\;,
\end{align}
while $\delta R^\lambda_{\;\;\alpha\beta\gamma}$ traces back to $\delta \Gamma^\lambda_{\alpha\beta}$ through  the Palatini identity
\begin{align}\label{Variation Step 4}
\delta R^\lambda_{\;\;\alpha\beta\gamma}\,=\,\nabla_\beta \big(\delta \Gamma^\lambda_{\gamma\alpha}\big)-\nabla_\gamma \big(\delta\Gamma^\lambda_{\beta\alpha}\big)\;.
\end{align}
Also, as is well known, $\delta \Gamma^\lambda_{\alpha\beta}\,=\,\frac{1}{2}\,g^{\lambda\sigma}\,\big(\nabla_\alpha \delta g_{\sigma\beta}+\nabla_\beta \delta g_{\sigma\alpha}-\nabla_\sigma \delta g_{\alpha\beta} \big)$\cite{DeWitt Specific Lagrangians}\cite{Hawking Ellis}, and we keep in mind that when raising the indices on $\delta g_{\alpha\beta}$ a  minus sign appears: $\delta g_{\alpha\beta}\,= -g_{\alpha\mu}\, g_{\beta\nu}\, \delta g^{\mu\nu}$. Then, Eqs.(\ref{Variation Step 2}-\ref{Variation Step 4}) yield
\begin{align}\label{Vary f(R)}
f_R\!\cdot\!\delta R\,\cong\,\bigg[f_R\,R_{\mu\nu}+\big(g_{\mu\nu}\Box-\nabla_\mu\!\nabla_\nu\big)\,f_R\bigg]\!\cdot\delta g^{\mu\nu}\,\eqqcolon H_{\mu\nu}^{(\!f{R})}\!\cdot\delta g^{\mu\nu}\;,
\end{align}
\begin{align}\label{Vary Rc2}
f_{R_c^2} \!\cdot \delta R_c^2 \cong \bigg[2f_{R_c^2}\,R_\mu^{\;\;\,\alpha}R_{\alpha\nu}
\!-\!\nabla_\alpha\!\nabla_{\nu} \big(R_{\mu}^{\;\;\,\alpha}f_{R_c^2}\big)\!-\!\nabla_\alpha\!\nabla_{\mu} \big(R_{\nu}^{\;\;\,\alpha}f_{R_c^2}\big)
\!+\!\Box \big(R_{\mu\nu}f_{R_c^2}\big) \!+\!g_{\mu\nu} \nabla_\alpha\!\nabla_\beta \big(R^{\alpha\beta} f_{R_c^2}  \big) \bigg]
\cdot \delta g^{\mu\nu}\:\eqqcolon H_{\mu\nu}^{(\!f{R_c^2})}\!\cdot \delta g^{\mu\nu}\,,
\end{align}
\begin{flalign}\label{Vary Rm2}%\hspace{1cm}
&\text{and }\hspace{2cm}f_{R_m^2}\!\cdot \delta R_m^2 \,\cong\,\bigg[2\,f_{R_m^2}\!\cdot\! R_{\mu\alpha\beta\gamma}R_{\nu}^{\;\;\,\alpha\beta\gamma}
+4\,\nabla^\beta \nabla^\alpha \big(R_{\alpha\mu \beta\nu}f_{R_m^2}  \big) \bigg]\!\cdot\delta g^{\mu\nu}
\;\eqqcolon H_{\mu\nu}^{(\!f{R_m^2})}\!\cdot\delta g^{\mu\nu}\; .&
\end{flalign}
Here, $\Box\equiv \nabla^\alpha \nabla_\alpha$ represents the covariant d'Alembertian, and the symbol $\cong$ denotes an effective equivalence by neglecting a surface integral after integration by parts twice to extract $\{H_{\mu\nu}^{(\!f{R})},H_{\mu\nu}^{(\!f{R_c^2})},H_{\mu\nu}^{(\!f{R_m^2})}\}$. Especially, Eq.(\ref{Vary Rm2}) has utilized the combination $2 \nabla^\beta \nabla^\alpha \big(R_{\alpha\mu \beta\nu}f_{R_m^2}  \big) \!+\!2 \nabla^\beta \nabla^\alpha \big(R_{\alpha\nu\beta\mu}f_{R_m^2}  \big)
 \!=\! 4 \nabla^\beta \nabla^\alpha \big(R_{\alpha\mu \beta\nu}f_{R_m^2}  \big)$, where the symmetry of $\nabla^\beta \nabla^\alpha \big(R_{\alpha\mu \beta\nu}f_{R_m^2} \big)$ under the index switch $\mu\leftrightarrow\nu$ is guaranteed by
$\nabla^\beta \nabla^\alpha  R_{\alpha\mu\beta\nu}$ $\!=\!\nabla^\beta \nabla^\alpha  R_{\alpha\mu\beta\nu}$ , $\nabla^\alpha\nabla^\beta f_{R_m^2}  \!=\! \nabla^\beta \nabla^\alpha f_{R_m^2}$
as well as the $\mu\leftrightarrow\nu$ symmetry of its remaining expanded terms.
Note that in these equations, total derivatives in individual variations $\{\delta R\,, \delta R_c^2\,, \delta R_m^2 \}$ are not necessarily
pure divergences anymore, because the nontrivial coefficients $\{f_R\,,f_{R_c^2}\,,f_{R_m^2}\}$ will be absorbed by the variations into the nonlinear and higher-order-derivative terms in $\{H_{\mu\nu}^{(\!f{R})},H_{\mu\nu}^{(\!f{R_c^2})},H_{\mu\nu}^{(\!f{R_m^2})}\}$.

In the $f_{\!\mathscr{L}_m}\!\cdot\delta\mathscr{L}_m$ term in Eq.(\ref{ActionVariation 1}),
we make use of the standard definition of  stress-energy-momentum (SEM) density tensor used in GR (e.g. \cite{Generalized Bianchi and Conservation 2}-\cite{f(R T)}),
which is introduced in accordance with minimal geometry-matter coupling and automatic energy-momentum conservation (for further discussion see Sec.~\ref{AA automati conservation minimal coupling}),
\begin{align}
T_{\mu\nu}\,\coloneqq& \frac{-2}{\sqrt{-g}}\frac{\delta(\!\sqrt{-g}\,\mathscr{L}_m)}{\delta g^{\mu\nu}}\label{Tmunu Definition} \\
=&\;\mathscr{L}_m\, g_{\mu\nu}-2\frac{\delta\mathscr{L}_m}{\delta g^{\mu\nu}} \;.\label{Tmunu Equivalence}
\end{align}
The equivalence from Eq.(\ref{Tmunu Definition}) to Eq.(\ref{Tmunu Equivalence}) is built upon the common assumption that $\mathscr{L}_m$ does not explicitly depend on derivatives of
the metric, $\mathscr{L}_m=\mathscr{L}_m(g_{\mu\nu},\psi_m)\neq \mathscr{L}_m(g_{\mu\nu},\partial_{\alpha}g_{\mu\nu},\psi_m)$ with $\psi_m$ collectively denoting all relevant matter fields.% 

After some work, Eqs.(\ref{ActionVariation 1}), (\ref{Vary f(R)}), (\ref{Vary Rc2}), (\ref{Vary Rm2}) and (\ref{Tmunu Equivalence}) eventually
give rise to the field equation for  $f(R,R_c^2,R_m^2, \mathscr{L}_m)$ gravity:
\begin{equation}\label{FieldEq-2}
-\frac{1}{2}f\,g_{\mu\nu}+f_R\,R_{\mu\nu}+\big(g_{\mu\nu}\Box-\nabla_\mu\!\nabla_\nu\big)\,f_R
+H_{\mu\nu}^{(\!fR_c^2)}+H_{\mu\nu}^{(\!fR_m^2)}\,=\,\frac{1}{2}\,f_{\!\mathscr{L}_m}\, \big(T_{\mu\nu}-\mathscr{L}_m\, g_{\mu\nu}\big)\;,
\end{equation}
where $H_{\mu\nu}^{(\!fR_c^2)}$ and $H_{\mu\nu}^{(\!fR_m^2)}$ were introduced in Eqs.(\ref{Vary Rc2}) and (\ref{Vary Rm2}) to collect all terms arising from $R_c^2$- and $R_m^2$-dependence in $f$,
\begin{equation}
\begin{split}
& H_{\mu\nu}^{(\!fR_c^2)}+H_{\mu\nu}^{(\!fR_m^2)}
 \;=2\, f_{R_c^2}\!\cdot R_\mu^{\;\;\,\alpha}R_{\alpha\nu}+2\, f_{R_m^2}\!\cdot\! R_{\mu\alpha\beta\gamma}R_{\nu}^{\;\;\,\alpha\beta\gamma}
 - \nabla_\alpha\!\nabla_{\nu} \big(R_{\mu}^{\;\;\,\alpha}f_{R_c^2}\big) \\
 &- \nabla_\alpha\!\nabla_{\mu} \big(R_{\nu}^{\;\;\,\alpha}f_{R_c^2}\big)
+  \Box \big(R_{\mu\nu}f_{R_c^2}\big)+ g_{\mu\nu}   \nabla_\alpha\!\nabla_\beta \big(R^{\alpha\beta} f_{R_c^2}  \big)
+4\, \nabla^\beta \nabla^\alpha \big(R_{\alpha\mu \beta\nu} f_{R_m^2}  \big) \;.\\
\end{split}
\label{Hmunu}
\end{equation}
Note that  $\{f,f_R,f_{R_c^2},f_{R_m^2}\}$ herein are all functions of $(R,R_c^2,R_m^2, \mathscr{L}_m)$, and $H_{\mu\nu}^{(\!fR)}\!=\!f_R R_{\mu\nu}\!+\!\big(g_{\mu\nu}\Box-\nabla_\mu\!\nabla_\nu\big) f_R$ has been written down directly to facilitate comparison with GR and $f(R)\!+\!2\kappa\mathscr{L}_m$ or $f(R,\mathscr{L}_m)$ gravity.
Taking the trace of Eq.(\ref{FieldEq-2}), the simple algebraic equality $R\!=\!-T$ (where $T\!=\!g^{\mu\nu}T_{\mu\nu}$) in GR is now generalized to the following differential relation,
\begin{equation}\label{FieldEq-Trace}
-2f+f_R\,R+2f_{R_c^2}\!\cdot R_c^2 +2\,f_{R_m^2}\!\cdot\! R_m^2+\Box\,\big(3 f_R  +f_{R_c^2}R\big)
+ 2 \nabla_\alpha\!\nabla_\beta \big(R^{\alpha\beta}f_{R_c^2}  +2R^{\alpha\beta}f_{R_m^2}   \big)
\,=\,f_{\!\mathscr{L}_m}\, \big(\frac{1}{2}T-2\mathscr{L}_m\big)\;.
\end{equation}

\vspace{1.6mm}Compared with  Einstein's equation $R_{\mu\nu}\!-\!Rg_{\mu\nu}/2\!=\!\kappa T_{\mu\nu}$ in GR, nonlinear terms
and derivatives of the metric up to fourth order have come forth and been encoded
into $\{H_{\mu\nu}^{(\!fR)}, H_{\mu\nu}^{(\!fR_c^2)}, H_{\mu\nu}^{(\!fR_m^2)}\}$ on the left hand side of Eq.(\ref{FieldEq-2}).
On the right hand side, the matter Lagrangian density $\mathscr{L}_m$ explicitly participates in the field equation as a consequence of
the confrontation between nonminimal curvature-matter coupling in $f(R,R_c^2,R_m^2, \mathscr{L}_m)$ and the minimal-coupling definition
of $T_{\mu\nu}$ in Eq.(\ref{Tmunu Definition}). Note that not all matter terms have been moved to the right hand side,
because  $-\frac{1}{2}fg_{\mu\nu}$ is still $\mathscr{L}_m$-dependent before a concrete $f(R,R_c^2,R_m^2, \mathscr{L}_m)$ model
gets specified and rearranged. % \\

Also, $f_{\!\mathscr{L}_m}=f_{\!\mathscr{L}_m}(R,R_c^2,R_m^2, \mathscr{L}_m)$ represents the
gravitational coupling strength and never vanishes, so  in vacuum one has $\mathscr{L}_m=0$ and $T_{\mu\nu}=0$ , yet $f_{\!\mathscr{L}_m}\neq0$.
Such a generic coupling strength  $f_{\!\mathscr{L}_m}$ will unavoidably violate Einstein's equivalence principle and the strong  equivalence principle unless it reduces to a constant.

%%%%%%%%%%%%%%%%%%%%%%%%%%%%%%%%%%%%%%%%%%%%%%%%%%%%%%%%%%%%%%%%%%%%%%%%%%%%%%%%%%%%%%%%%%%%
%%%%%%%%%%%%%%%%%%%%%%%%%%%%%%%%%%%%%%%%%%%%%%%%%%%%%%%%%%%%%%%%%%%%%%%%%%%%%%%%%%%%%%%%%%%%

\subsection{Field equation under minimal coupling}

When the matter content is minimally coupled to the spacetime metric, the coupling coefficient  $f_{\!\mathscr{L}_m}$ reduces to become a constant.
In accordance with the gravitational coupling strength in GR, this constant is necessarily equal to Einstein's constant $\kappa$
(and doubled just for scaling tradition). That is,
\begin{equation}\label{minimal coupling}
f_{\!\mathscr{L}_m}=\text{constant}\,=\,2\kappa\quad,\quad f(R\,,R_c^2\,,R_m^2\,, \mathscr{L}_m)\,=\,\tilde{f}(R\,,R_c^2\,,R_m^2)+2\kappa\,\mathscr{L}_m\;.
\end{equation}
We have neglected the situation when $f_{\!\mathscr{L}_m}$ is a pointwise scalar field $\phi=\phi(x^\alpha)$, which should be treated
as a scalar-tensor theory mixed with metric gravity: in fact, $\phi(x^\alpha)\,\mathscr{L}_m$
is also a type of nonminimal coupling, but it goes beyond the scope of this paper and will not be discussed here.
Under minimal coupling as in Eq.(\ref{minimal coupling}), the field equation  (\ref{FieldEq-2})
becomes (with tildes on $\tilde{f}$ omitted)
\begin{equation}
-\frac{1}{2}f\,g_{\mu\nu}+f_R\,R_{\mu\nu}+\big(g_{\mu\nu}\Box-\nabla_\mu\!\nabla_\nu\big)\,f_R
+H_{\mu\nu}^{(\!fR_c^2)}+H_{\mu\nu}^{(\!fR_m^2)}\,=\,\kappa\,T_{\mu\nu}\;,
\end{equation}
which coincides with the result in \cite{Carroll R+ f(R Rc2 Rm2)+2kLm}. The weak field limit of this minimally coupled model has been systematically studied in \cite{WeakFieldLimit-2}.

%%%%%%%%%%%%%%%%%%%%%%%%%%%%%%%%%%%%%%%%%%%%%%%%%%%%%%%%%%%%%%%%%%%%%%%%%%%%%%%%%%%%%
%%%%%%%%%%%%%%%%%%%%%%%%%%%%%%%%%%%%%%%%%%%%%%%%%%%%%%%%%%%%%%%%%%%%%%%%%%%%%%%%%%%%%

\subsection{Two types of nonminimal curvature-matter coupling}
Apart from the $\mathscr{L}\!=\!f(R,R_c^2,R_m^2,\mathscr{L}_m)$ model under discussion, another type of curvature-matter coupling
was introduced in \cite{f(R T)} by the $\mathscr{L}\!=\!f(R,T)+2\kappa \mathscr{L}_m$ model, where
a curvature invariant was nonminimally coupled to the trace of the SEM tensor $T\!=\!g^{\mu\nu}T_{\mu\nu}$ rather than the matter
Lagrangian density $\mathscr{L}_m$. In this spirit, we consider the following nonminimally  coupled action,
\begin{equation}\label{T nonminimal coupling Action}
\mathcal{S}\,=\,\int d^4x \sqrt{-g}\,\bigg\{f(R\,,R_c^2\,,R_m^2\,,T)+2\kappa\, \mathscr{L}_m\bigg\}\;.
\end{equation}
By the standard methods we find that its field equation is:
\begin{equation}\label{T nonminimal coupling Field Equation}
%\begin{split}
-\frac{1}{2}f\,g_{\mu\nu}+f_R\!\cdot\! R_{\mu\nu}+\big(g_{\mu\nu}\Box-\nabla_\mu\!\nabla_\nu\big)\,f_R +H_{\mu\nu}^{(\!fR_c^2)}+H_{\mu\nu}^{(\!fR_m^2)}\,
=-f_T\!\cdot\!  \big( T_{\mu\nu}+  \Theta_{\mu\nu} \big)+\kappa T_{\mu\nu}\;,
%\end{split}
\end{equation}
where $\{f,f_R,f_{R_c^2},f_T\}$ are all functions of $(R,R_c^2,R_m^2, T)$ , $H_{\mu\nu}^{(\!fR_c^2)}+H_{\mu\nu}^{(\!fR_m^2)}$ is given by Eq.(\ref{Hmunu}), $-f_T\, \big( T_{\mu\nu}+  \Theta_{\mu\nu} \big)$ comes from the $T$-dependence in
$f(R,R_c^2,R_m^2,T)$, and
\begin{equation}\label{T Theta munu}
\Theta_{\mu\nu}\,\coloneqq\, \frac{g^{\alpha\beta}\,\delta T_{\alpha\beta}}{\delta g^{\mu\nu}} \;.
\end{equation}
{As will be extensively discussed in Section 5, for some matter sources  $\mathscr{L}_m$ cannot be uniquely specified, and therefore the equations of continuity and motion based on Eq.(\ref{FieldEq-2}) have to rely on the choice of  $\mathscr{L}_m$. In such situations $T_{\mu\nu}$ is easier to set up than $\mathscr{L}_m$,
so at first glance, it seems as if the new field equation ($\ref{T nonminimal coupling Field Equation}$) could avoid the flaws
from nonminimal $\mathscr{L}_m$-coupling,
at the cost of employing a supplementary matter tensor $\Theta_{\mu\nu}$.} However, the definition of $\Theta_{\mu\nu}$ is still
based on the relation $T_{\mu\nu}=\mathscr{L}_m g_{\mu\nu}-2\delta\mathscr{L}_m/\delta g^{\mu\nu}$ in Eq.(\ref{Tmunu Equivalence}), and explicit
calculations have revealed that\cite{f(R T)}
\begin{equation}
\Theta_{\mu\nu}\,=-2T_{\mu\nu}+g_{\mu\nu}\mathscr{L}_m-2g^{\alpha\beta}\,\frac{\partial^2\mathscr{L}_m}{\partial g^{\mu\nu}\partial g^{\alpha\beta}}\;.
\end{equation}
Thus, both  $\mathscr{L}_m$ and its second-order derivative with respect to the metric are hidden in $\Theta_{\mu\nu}$, and consequently, both
$f(R,R_c^2,R_m^2,T)+2\kappa\mathscr{L}_m$ and $f(R,R_c^2,R_m^2,\mathscr{L}_m)$ theories {are sensitive to the $\mathscr{L}_m$ in use}.
The equations of continuity and nongeodesic motion will differ for different choices of $\mathscr{L}_m$ for the same matter source. % \\

The  $\mathscr{L}\!=\!f(R,R_c^2,R_m^2,\mathscr{L}_m)$ model and the $\mathscr{L}\!=\!f(R,R_c^2,R_m^2,T)+2\kappa\mathscr{L}_m$ model are
both reasonable realizations of nonminimal
curvature-matter coupling, and in this paper we have adopted the former case as a generalization of the existing $\mathscr{L}\!=\!f(R,\mathscr{L}_m)$\cite{f(R Lm)} and
$\mathscr{L}\!=\!f(R,R_c^2,R_m^2)+2\kappa\mathscr{L}_m$\cite{Carroll R+ f(R Rc2 Rm2)+2kLm} theories. Also, it looks redundant and unnecessary to further consider the superposition of
nonminimal $\mathscr{L}_m$-  and $T$-couplings, which can be depicted by the action
\begin{equation}\label{Nonminimal Lm T couplings superposition}
\mathcal{S}\,=\,\int d^4x \sqrt{-g}\, f(R\,,R_c^2\,,R_m^2\,,\mathscr{L}_m\,,T) \;,
\end{equation}
whose field equation is
\begin{equation}
%\begin{split}
-\frac{1}{2}f\,g_{\mu\nu}+f_R\!\cdot\! R_{\mu\nu}+\big(g_{\mu\nu}\Box-\nabla_\mu\!\nabla_\nu\big)\,f_R +H_{\mu\nu}^{(\!fR_c^2)}+H_{\mu\nu}^{(\!fR_m^2)}\,
=\,\frac{1}{2}f_{\!\mathscr{L}_m}\!\cdot  \big(T_{\mu\nu}-\mathscr{L}_m\, g_{\mu\nu}\big)-f_T\!\cdot\!  \big( T_{\mu\nu}+  \Theta_{\mu\nu} \big)\;.
%\end{split}
\end{equation}
Practically it is implicitly assumed in Eq.(\ref{Nonminimal Lm T couplings superposition}) that nonminimal couplings happen between $(R\,,R_c^2\,,R_m^2\,,\mathscr{L}_m)$ and $(R\,,R_c^2\,,R_m^2\,,T)$ respectively, and there is no matter-matter  $\mathscr{L}_m$-$T$ coupling which would cause severe theoretical complexity and physical ambiguity. In fact,
$\mathscr{L}_m$ and $T$ are not independent, as Eq.(\ref{Tmunu Equivalence}) implies that
\begin{equation}
T\,=\,g^{\alpha\beta}T_{\alpha\beta}\,%=\,g^{\alpha\beta}\,(\mathscr{L}_m\, g_{\alpha\beta}-2\frac{\delta\mathscr{L}_m}{\delta g^{\alpha\beta}} )
=\,4\mathscr{L}_m-2g^{\alpha\beta}\,\frac{\delta\mathscr{L}_m}{\delta g^{\alpha\beta}}\;.
\end{equation}

%%%%%%%%%%%%%%%%%%%%%%%%%%%%%%%%%%%%%%%%%%%%%%%%%%%%%%%%%%%%%%%%%%%%%%%%%%%%%%%%%%%%%%%%%%%%
%%%%%%%%%%%%%%%%%%%%%%%%%%%%%%%%%%%%%%%%%%%%%%%%%%%%%%%%%%%%%%%%%%%%%%%%%%%%%%%%%%%%%%%%%%%%

\section{$R^2$-dependence, smooth transition to generalized Gauss-Bonnet gravity, and quadratic gravity}\label{Section 3}

Generalized (Einstein-)Gauss-Bonnet gravity is perhaps the most popular and typical situation in which there is dependence on $R$ and the quadratic invariants $\{R_c^2,R_m^2\}$\cite{GaussBonnet first model R/2k+f(G)}\cite{GaussBonnet second model f(R G)+Lm}. However, to the best of our knowledge, there is no
demonstration of how generic fourth-order model $f(R,R_c^2,R_m^2,\mathscr{L}_m)$ (or $f(R,R_c^2,R_m^2)+2\kappa \mathscr{L}_m$ model if minimally coupled\cite{Carroll R+ f(R Rc2 Rm2)+2kLm}) may be smoothly reduced into generalized Gauss-Bonnet theories. We tackle this problem by considering an explicit dependence on $R^2$ in $f(R,R_c^2,R_m^2,\mathscr{L}_m)$ gravity.

\subsection{Two generic $R^2$-dependent models}
Based on the $f(R,R_c^2,R_m^2,\mathscr{L}_m)$ gravity, we consider the following   situation with an explicit
dependence on $R^2$:
\begin{equation}\label{Action with explicit R2}
\mathscr{L}\,=\,f(R\,,R^2\,,R_c^2\,,R_m^2\,,\mathscr{L}_m)\;.
\end{equation}
Here we have formally split the generic $R$-dependence of $f(R,R_c^2,R_m^2,\mathscr{L}_m)$ into an $R$- and $R^2$-dependence, $f_R \,\delta R\mapsto f_R  \,\delta R+f_{R^2} \, \delta R^2$, to lay the foundation for subsequent discussion. However, this $f(R,R^2,R_c^2,R_m^2,\mathscr{L}_m)$ Lagrangian density is not more generic than $f(R,R_c^2,R_m^2,\mathscr{L}_m)$ by one more variable $R^2$.
Absorbing $f_{R^2}$ into $\delta R^2\!=\!2R  \,\delta R$ by the replacement $f_R\mapsto 2R\, f_{R^2}$ in Eq.(\ref{Vary f(R)}), we learn that $R^2$-dependence would contribute to the field equation by
\begin{equation}\label{Vary R2}
f_{R^2}\!\cdot\!\delta R^2\,\cong\,\bigg[2R\,f_{R^2}\!\cdot\!R_{\mu\nu}
+2\,\big(g_{\mu\nu}\Box-\nabla_\mu\!\nabla_\nu\big)\,\big(R\!\cdot\!f_{R^2}\big)\bigg]\!\cdot\!\delta g^{\mu\nu}\;\eqqcolon H_{\mu\nu}^{(\!fR^2)} \!\cdot \delta g^{\mu\nu}\;,
\end{equation}
and a resubstitution of $f_R \!\mapsto\! f_R\!+\!2Rf_{R^2}$ into Eq.(\ref{FieldEq-2})  directly yields the field equation for $f(R,R^2,R_c^2,R_m^2,\mathscr{L}_m)$ gravity,
\begin{equation}\label{FieldEq Explicit R2}
-\frac{1}{2}f\,g_{\mu\nu}+f_R\,R_{\mu\nu}+\big(g_{\mu\nu}\Box-\nabla_\mu\!\nabla_\nu\big)\,
f_R+H_{\mu\nu}^{(\!fR^2)}+H_{\mu\nu}^{(\!fR_c^2)}+H_{\mu\nu}^{(\!fR_m^2)} \,=\,\frac{1}{2}f_{\!\mathscr{L}_m}\, \big(T_{\mu\nu}-\mathscr{L}_m\, g_{\mu\nu}\big)\;,
\end{equation}
where $\{f, f_R, f_{R^2}\}$ and the $\{ f_{R_c^2}, f_{R_m^2}\}$ in $\{H_{\mu\nu}^{(\!fR_c^2)}+ H_{\mu\nu}^{(\!fR_m^2)}\}$ are all functions of $(R,R^2,R_c^2,R_m^2, \mathscr{L}_m)$.% \\

Here we have assumed no ambiguity between the $R$-dependence and the $R^2$-dependence in Eq.(\ref{Action with explicit R2}). To explicitly avoid this problem,  one could consider a Lagrangian density of the form,
\begin{equation}\label{Action with explicit R2 II}
\mathscr{L}\,=\,\tilde{f}(R)+f(R^2\,,R_c^2\,,R_m^2\,,\mathscr{L}_m)\;.
\end{equation}
%so as to avoid ambiguity by formally putting the $R$- and $R^2$-dependence into two separate functions $\tilde{f}$ and $f$, $f_R \,\delta R\mapsto \tilde{f}_R  \,\delta %R+f_{R^2} \, \delta R^2$.
However, potential coupling between $R^2$ and $\mathscr{L}_m$ can still be turned around and retreated as $R-\mathscr{L}_m$ coupling, so this $\tilde{f}(R)+f(R^2,R_c^2,R_m^2,\mathscr{L}_m)$ model is still equally generic with $f(R,R_c^2,R_m^2,\mathscr{L}_m)$
as well as the $f(R,R^2,R_c^2,R_m^2,\mathscr{L}_m)$ just above. Setting $f\mapsto \tilde{f}+f$ and $f_R\mapsto \tilde{f}_R+2Rf_{R^2}$ in Eq.(\ref{FieldEq-2}), we get the field equation for Eq.(\ref{Action with explicit R2 II}),
\begin{equation}\label{FieldEq Explicit R2 II}
\begin{split}
-\frac{1}{2}\big(\tilde{f}+f\big)\,g_{\mu\nu}+\tilde{f}_R\,R_{\mu\nu}
+\big(g_{\mu\nu}\Box-\nabla_\mu\!\nabla_\nu\big)\,\tilde{f}_R+H_{\mu\nu}^{(\!fR^2)}+H_{\mu\nu}^{(\!fR_c^2)}+H_{\mu\nu}^{(\!fR_m^2)}  \,=\,\frac{1}{2}f_{\!\mathscr{L}_m}\, \big(T_{\mu\nu}-\mathscr{L}_m\, g_{\mu\nu}\big)&\;,
\end{split}
\end{equation}
where $\tilde{f}_R=\tilde{f}_R(R)$, $f_{R^2}=f_{R^2}(R^2,R_c^2,R_m^2,\mathscr{L}_m)$, and $\{ f_{R_c^2}, f_{R_m^2}\}$  remain dependent on $(R,R^2,R_c^2,R_m^2, \mathscr{L}_m)$. Moreover, Eq.(\ref{FieldEq Explicit R2 II}) can instead be obtained from Eq.(\ref{FieldEq Explicit R2}) by the replacement $f_R \mapsto \tilde{f}_R$.% \\

For subsequent investigations, it will be sufficient to just employ the former model
$\mathscr{L}\!=\!f(R,R^2,R_c^2,R_m^2,\mathscr{L}_m)$ and its field equation (\ref{FieldEq Explicit R2}).

%%%%%%%%%%%%%%%%%%%%%%%%%%%%%%%%%%%%%%%%%%%%%%%%%%%%%%%%%%%%%%%%%%%%%%%%%%%%%%%%%%%%%%%%%%%%
%%%%%%%%%%%%%%%%%%%%%%%%%%%%%%%%%%%%%%%%%%%%%%%%%%%%%%%%%%%%%%%%%%%%%%%%%%%%%%%%%%%%%%%%%%%%

\subsection{Reduced field equation with $f_{R^2}=f_{R_m^2}=  -f_{R_c^2}/4$}

Now recall that the second Bianchi identity $\nabla_\gamma R_{\alpha\mu\beta\nu}+\nabla_\nu R_{\alpha\mu\gamma\beta}+\nabla_\beta R_{\alpha\mu\nu\gamma}=0$ implies the following simplifications,
which rewrite the derivative of a high-rank curvature tensor into that of lower-rank curvature tensors plus nonlinear algebraic terms:
\begin{align}
\nabla^\alpha R_{\alpha\mu\beta\nu}\,&=\,\nabla_\beta R_{\mu\nu}-\nabla_\nu R_{\mu\beta}\label{Bianchi implications 1} \\
\nabla^\alpha R_{\alpha\beta} \,&=\,\frac{1}{2}\,\nabla_\beta R \label{Bianchi implications 2} \\
\nabla^\beta\nabla^\alpha  R_{\alpha\beta}\,&=\,\frac{1}{2}\,\Box R \label{Bianchi implications 3}\\
\nabla^\beta \nabla^\alpha  R_{\alpha\mu\beta\nu}\,=\,\Box R_{\mu\nu}\,&-\,\frac{1}{2}\nabla_\mu\! \nabla_\nu R+R_{\alpha\mu\beta\nu}R^{\alpha\beta}-R_{\mu}^{\;\;\,\alpha}R_{\alpha\nu}\label{Bianchi implications 4}\\
\vspace{5mm}\nabla^\alpha\nabla_\mu R_{\alpha\nu}+\nabla^\alpha\nabla_\nu R_{\alpha\mu}\,&=\, \nabla_\mu\!\nabla_\nu R -2R_{\alpha\mu\beta\nu}R^{\alpha\beta}+2R_{\mu}^{\;\;\,\alpha}R_{\alpha\nu}\;,\label{Bianchi implications 5}
\end{align}
along with the symmetry $\nabla^\beta \nabla^\alpha  R_{\alpha\mu\beta\nu}=\nabla^\beta \nabla^\alpha  R_{\alpha\nu\beta\mu}$
and $\nabla^\alpha\nabla_\mu R_{\alpha\nu}+\nabla^\alpha\nabla_\nu R_{\alpha\mu}=2\,\big(\Box R_{\mu\nu}
-\nabla^\beta \nabla^\alpha  R_{\alpha\mu\beta\nu}\big)$. Applying these relations to expand
all the second-order covariant derivatives in Eq.(\ref{FieldEq Explicit R2}), it turns out that we have the following theorem.\\

\noindent \emph{Theorem}: When the coefficients $\{f_{R^2}\,, f_{R_c^2}\,,f_{R_m^2}  \}$ satisfy the following proportionality conditions,
\begin{equation}\label{Coherence Takeout Condition}
f_{R^2}\,=\,f_{R_m^2}\,=  -\frac{1}{4}f_{R_c^2}\;\eqqcolon\,F\;,
\end{equation}
where $F=F(R,R^2,R_c^2,R_m^2,\mathscr{L}_m)$, then the field equation (\ref{FieldEq Explicit R2}) reduces to
\begin{equation}\label{FieldEq Explicit R2 Simplified using F}
-\frac{1}{2}f\,g_{\mu\nu}+f_R\,R_{\mu\nu}+\big(g_{\mu\nu}\Box-\nabla_\mu\!\nabla_\nu\big)\,f_R +\mathcal{H}_{\mu\nu}^{(F)} \,=\,\frac{1}{2}f_{\!\mathscr{L}_m}\, \big(T_{\mu\nu}-\mathscr{L}_m\, g_{\mu\nu}\big)\;,
\end{equation}
where
\begin{equation}\label{Hmunu Coherence}
\begin{split}
\mathcal{H}_{\mu\nu}^{(F)} \,\coloneqq\:\;& 2Rf_{R^2}\!\cdot\! R_{\mu\nu}-4f_{R_m^2}\!\cdot\! R_\mu^{\;\;\,\alpha}R_{\alpha\nu} + \big(2f_{R_c^2}+4 f_{R_m^2}\big)\!\cdot\!R_{\alpha\mu\beta\nu}R^{\alpha\beta}
+2f_{R_m^2}\!\cdot\! R_{\mu\alpha\beta\gamma}R_{\nu}^{\;\;\,\alpha\beta\gamma}\\
&+2R\big(g_{\mu\nu}\Box
-\nabla_\mu\!\nabla_\nu\big)\,f_{R^2}-R_{\mu}^{\;\;\,\alpha} \nabla_\alpha\!\nabla_{\nu}f_{R_c^2}
-R_{\nu}^{\;\;\,\alpha}\nabla_\alpha\!\nabla_{\mu} f_{R_c^2} +R_{\mu\nu}\Box f_{R_c^2} \\
&+g_{\mu\nu} \!\cdot\! R^{\alpha\beta}\nabla_\alpha\!\nabla_\beta f_{R_c^2} +4\,R_{\alpha\mu \beta\nu} \nabla^\beta \nabla^\alpha  f_{R_m^2}\;\;(f_{R^2}=f_{R_m^2}=  -f_{R_c^2}/4 )\\
\equiv\:\;& 2RF\!\cdot\! R_{\mu\nu}-4F\!\cdot\! R_\mu^{\;\;\,\alpha}R_{\alpha\nu}-4F\!\cdot\!R_{\alpha\mu\beta\nu}R^{\alpha\beta} +2F\!\cdot\! R_{\mu\alpha\beta\gamma}R_{\nu}^{\;\;\,\alpha\beta\gamma}\\
&+2R\big(g_{\mu\nu}\Box
-\nabla_\mu\!\nabla_\nu\big)\,F +4R_{\mu}^{\;\;\,\alpha}\nabla_\alpha\!\nabla_{\nu}F +4R_{\nu}^{\;\;\,\alpha}\nabla_\alpha\!\nabla_{\mu} F\\
 &-4R_{\mu\nu}\Box F -4g_{\mu\nu} \!\cdot\! R^{\alpha\beta}\nabla_\alpha\!\nabla_\beta F+4\,R_{\alpha\mu \beta\nu} \nabla^\beta \nabla^\alpha  F\;.
\end{split}
\end{equation}
$\mathcal{H}_{\mu\nu}^{(F)}\delta g^{\mu\nu}= f_{F}\,\delta F$ and second-order-derivative operators $\{\Box,\nabla_\alpha\!\nabla_\nu, \text{etc}\}$  only act on the scalar functions $\{f_{R^2}\,, f_{R_c^2}\,,f_{R_m^2}  \}$ in contrast to $H_{\mu\nu}^{(\!fR^2)}\!+\! H_{\mu\nu}^{(\!fR_c^2)}\!+\!H_{\mu\nu}^{(\!fR_m^2)} $ in Eq.(\ref{Action with explicit R2})\footnote{This is also why we use the denotation $\mathcal{H}_{\mu\nu}^{(F)}$ rather than $H_{\mu\nu}^{(F)}$}. \\

Note that similar techniques have been employed in \cite{GaussBonnet mixed with dynamical scalar field} to finalize the field equation of the dilaton-Gauss-Bonnet model. The simplified field equation (\ref{FieldEq Explicit R2 Simplified using F}) after imposing the proportionality condition Eq.(\ref{Coherence Takeout Condition}) to Eq.(\ref{FieldEq Explicit R2}) will serve as a bridge connecting $f(R,R^2,R_c^2,R_m^2,\mathscr{L}_m)$ gravity to generalized Gauss-Bonnet gravity.  We refer to the proportionality condition Eq.(\ref{Coherence Takeout Condition}) as the \emph{coherence condition} to highlight the fact that it aligns the behaviors of $\{f_{R^2}, f_{R_c^2}, f_{R_m^2}\}$, and call $F$ therein the \emph{coherence function}.

\subsection{Generalized Gauss-Bonnet gravity with nonminimal coupling}
\subsubsection{Generic $\mathscr{L}=f(R,\mathcal{G}, \mathscr{L}_m)$ model}
A nice way to realize the coherence condition Eq.(\ref{Coherence Takeout Condition})  is to let  $\{R^2,R_c^2,R_m^2\}$ participate in the action
through the well-known  Gauss-Bonnet invariant $\mathcal G$,
\begin{equation}
\mathcal G\,\coloneqq\,R^2-4R_c^2+R_m^2\;.
\end{equation}
In this case, Eq.(\ref{Action with explicit R2}) reduces to become the Lagrangian density of a generalized Gauss-Bonnet gravity model allowing nonminimal curvature-matter coupling,
\begin{equation}\label{Gauss Bonnet most general action nonminimal coupling}
\mathscr{L}\,=\, f(R\,,\mathcal{G}\,,\mathscr{L}_m)\;.
\end{equation}
Then the proportionality in Eq.(\ref{Coherence Takeout Condition}) is naturally satisfied with the coherence function $F$
recognized as $f_{\mathcal{G}}\coloneqq\partial f/\partial \mathcal{G}$. Given $F \mapsto f_{\mathcal{G}}$,
 Eqs.(\ref{Hmunu Coherence}) and (\ref{FieldEq Explicit R2 Simplified using F}) give rise to the field equation for  $f(R,\mathcal{G},\mathscr{L}_m)$ gravity right away,
\begin{equation}\label{Gauss-Bonnet Most Generic Field Eqn}
-\frac{1}{2}f\,g_{\mu\nu}+ f_R\,R_{\mu\nu}
+\big(g_{\mu\nu}\Box-\nabla_\mu\!\nabla_\nu\big)\,f_R
+\mathcal{H}_{\mu\nu}^{\text{(GB)}}\,=\,\frac{1}{2}\,f_{\!\mathscr{L}_m}\, \big(T_{\mu\nu}-\mathscr{L}_m\, g_{\mu\nu}\big) \;,
\end{equation}
where
\begin{equation}\label{Gauss-Bonnet Most Generic Hmunu}
\begin{split}
\mathcal{H}_{\mu\nu}^{\text{(GB)}}\coloneqq  2f_{\mathcal{G}}\!\cdot\! RR_{\mu\nu} -4f_{\mathcal{G}}\!\cdot R_\mu^{\;\;\,\alpha}R_{\alpha\nu}\!-\!4f_{\mathcal{G}}\!\cdot\!R_{\alpha\mu\beta\nu}R^{\alpha\beta}
\!+\!2f_{\mathcal{G}}\!\cdot\! R_{\mu\alpha\beta\gamma}R_{\nu}^{\;\;\,\alpha\beta\gamma}
 +2R\,\big(g_{\mu\nu}\Box-\nabla_\mu\!\nabla_\nu\big)\,f_{\mathcal{G}} &\\
+\, 4R_{\mu}^{\;\;\,\alpha}\nabla_\alpha\!\nabla_{\nu}f_{\mathcal{G}} +\,4R_{\nu}^{\;\;\,\alpha}\nabla_\alpha\!\nabla_{\mu} f_{\mathcal{G}}  -\,4R_{\mu\nu}\Box f_{\mathcal{G}} -\,4g_{\mu\nu} \!\cdot\! R^{\alpha\beta}\nabla_\alpha\!\nabla_\beta f_{\mathcal{G}}+\,4\,R_{\alpha\mu \beta\nu} \nabla^\beta \nabla^\alpha  f_{\mathcal{G}}&\;,
\end{split}
\end{equation}
and $\{f,f_R,f_{\mathcal{G}}\}$ are all functions of $(R,\mathcal{G},\mathscr{L}_m)$ , and $\mathcal{H}_{\mu\nu}^{\text{(GB)}}\delta g^{\mu\nu}= f_{\mathcal{G}}\,\delta \mathcal{G}$.

%%%%%%%%%%%%%%%%%%%%%%%%%%%%%%%%%%%%%%%%%%%%%%%%%%%%%%%%%%%%%%%%%%%%%%%%%%%%%%%%%%%
%%%%%%%%%%%%%%%%%%%%%%%%%%%%%%%%%%%%%%%%%%%%%%%%%%%%%%%%%%%%%%%%%%%%%%%%%%%%%%%%%%%

\subsubsection{No contributions from  a pure Gauss-Bonnet term}

As for the $\mathcal{G}$-dependence,
Eqs.(\ref{Gauss-Bonnet Most Generic Field Eqn}) and (\ref{Gauss-Bonnet Most Generic Hmunu}) are best simplified
when $f_{\mathcal{G}}\!=\!\lambda\!=$constant; that is to say, $\mathcal{G}$ joins $\mathscr{L}$ straightforwardly as a pure Gauss-Bonnet term,
with Lagrangian density $\mathscr{L}\!=\!f(R,\mathscr{L}_m)+\lambda\mathcal{G}$, for which Eq.(\ref{Gauss-Bonnet Most Generic Field Eqn}) gives rise to the field equation (with $f=f(R,\mathscr{L}_m)$, $f_R=f_R(R,\mathscr{L}_m)$):
\begin{equation}\label{Gauss-Bonnet Best Simplified}
\begin{split}
\lambda\!\cdot\!\Big(-\frac{1}{2}\mathcal{G}\,g_{\mu\nu}+2 R\,R_{\mu\nu}-4 R_\mu^{\;\;\,\alpha}R_{\alpha\nu}-4 R_{\alpha\mu\beta\nu}R^{\alpha\beta}
+2R_{\mu\alpha\beta\gamma}R_{\nu}^{\;\;\,\alpha\beta\gamma}\Big)&\\
-\frac{1}{2}f\,g_{\mu\nu}+f_R\,R_{\mu\nu}+\big(g_{\mu\nu}\Box-\nabla_\mu\!\nabla_\nu\big)\,f_R
=\,\frac{1}{2}f_{\!\mathscr{L}_m}\, \big(T_{\mu\nu}-\mathscr{L}_m\, g_{\mu\nu}\big)&\;.
\end{split}
\end{equation}

\vspace{2mm}At first glance, it may seem that, after $\mathcal{G}$ decouples from $f(R,\mathcal{G}, \mathscr{L}_m)$  to form a pure term $\lambda\,\mathcal{G}$, the isolated covariant density  $\lambda\!\sqrt{-g}\,\mathcal{G}$ would still make a difference to the field equation by the $\lambda\!\cdot\!\big(\ldots\big)$ term in Eq.(\ref{Gauss-Bonnet Best Simplified}).  This result conflicts our \emph{a priori} anticipation that, since  $\mathcal{G}$ is a topological invariant,  variation of the Euler-Poincar\'e topological density $\sqrt{-g}\,\mathcal G$  should not change the gravitational field equation. In fact, by setting $f_{R^2}=f_{R_c^2}=f_{R_m^2}=1$ in
Eqs.(\ref{Vary Rc2}), (\ref{Vary Rm2}) and (\ref{Vary R2}), one has
\begin{equation}
\delta R^2/\delta g^{\mu\nu}\,=\, 2R\,R_{\mu\nu}+2\,\big(g_{\mu\nu}\Box-\nabla_\mu\!\nabla_\nu\big)\,R \;,
\end{equation}
\begin{equation}
\hspace{1cm}\delta R_c^2/\delta g^{\mu\nu} \,= \, 2 R_\mu^{\;\;\,\alpha}R_{\alpha\nu} -\nabla_\alpha\!\nabla_{\nu} R_{\mu}^{\;\;\,\alpha}-\nabla_\alpha\!\nabla_{\mu} R_{\nu}^{\;\;\,\alpha}
+\Box R_{\mu\nu} +g_{\mu\nu} \!\cdot\! \nabla_\alpha\!\nabla_\beta R^{\alpha\beta}\;,\;\;\text{and}
\end{equation}
\begin{equation}
\delta R_m^2/\delta g^{\mu\nu} \,=\,2\,R_{\mu\alpha\beta\gamma}R_{\nu}^{\;\;\,\alpha\beta\gamma}+4\,\nabla^\beta \nabla^\alpha R_{\alpha\mu \beta\nu}  \;,
\end{equation}
which together with the Bianchi implications Eqs.(\ref{Bianchi implications 1})-(\ref{Bianchi implications 5}) exactly lead to
\begin{equation}\label{Gauss-Bonnet isolated variation}
\delta\, \big(\!\sqrt{-g}\,\mathcal{G}\big)/\delta g^{\mu\nu}
\,=\,-\frac{1}{2}\mathcal{G}\,g_{\mu\nu}+2 R\,R_{\mu\nu}-4 R_\mu^{\;\;\,\alpha}R_{\alpha\nu}-4 R_{\alpha\mu\beta\nu}R^{\alpha\beta}
+2R_{\mu\alpha\beta\gamma}R_{\nu}^{\;\;\,\alpha\beta\gamma}\;.
\end{equation}
Thus one can recover the term $\lambda\!\cdot\!\big(\ldots\big)$ in Eq.(\ref{Gauss-Bonnet Best Simplified}) by directly varying the quadratic invariants comprising $\mathcal{G}$.% \\

However, in four dimensions $\mathcal{G}$ is a most special invariant among all algebraic and differential Riemannian invariants $\mathcal{R} = \mathcal{R}(g_{\alpha\beta} ,R_{\alpha\mu\beta\nu} ,\nabla_\gamma R_{\alpha\mu\beta\nu} ,
\ldots ,\nabla_{\gamma_ 1}\!\nabla_{\gamma_ 2}\ldots \nabla_{\gamma_ n} R_{\alpha\mu\beta\nu})$ in the sense that it respects the Bach-Lanczos identity
\begin{equation}\label{Gauss-Bonnet Bach-Lanczos identity}
\delta\int dx^4 \sqrt{-g}\,\mathcal{G} \,\equiv\,0\;,
\end{equation}
which prevents the Gauss-Bonnet covariant density $\lambda\!\sqrt{-g}\,\mathcal{G}$ from contributing to the field equation. This identity can be verified by carrying out the variational derivative\cite{DeWitt Specific Lagrangians}\cite{Generalized Bianchi Lovelock and Rund Books}
\begin{equation}
\frac{\delta \big(\!\!\sqrt{-g}\,\mathcal{G}\big)}{\delta g^{\mu\nu}}\,=\,
\frac{ \partial \big(\!\!\sqrt{-g}\,\mathcal{G}\big)}{\partial g^{\mu\nu}}-\partial_\alpha \frac{ \partial \big(\!\!\sqrt{-g}\,\mathcal{G}\big)}{\partial (\partial_\alpha g^{\mu\nu})} + \partial_\alpha \partial_\beta\, \frac{\partial \big(\!\!\sqrt{-g}\,\mathcal{G}\big)}{\partial (\partial_\alpha \partial_\beta g^{\mu\nu})}\,\equiv\,0\;.
\end{equation}
On the other hand, algebraic identities satisfied by the Riemann tensor also guarantee that $-\frac{1}{2}\mathcal{G}\,g_{\mu\nu}+2 R\,R_{\mu\nu}-4 R_\mu^{\;\;\,\alpha}R_{\alpha\nu}-4 R_{\alpha\mu\beta\nu}R^{\alpha\beta}
+2R_{\mu\alpha\beta\gamma}R_{\nu}^{\;\;\,\alpha\beta\gamma}=0$\cite{DeWitt Specific Lagrangians}. % \\

Hence, the $\lambda\!\cdot \big(\ldots\big)$ term in Eq.(\ref{Gauss-Bonnet Best Simplified}), as a remnant of degrading the generic $f(R,\mathcal{G}, \mathscr{L}_m)$  gravity and all existing generalized Gauss-Bonnet theories, is removable,  and Eq.(\ref{Gauss-Bonnet Best Simplified})  for  $\mathscr{L}=f(R,\mathscr{L}_m)+\lambda\mathcal{G}$  gravity  finally becomes
\begin{equation}\label{Gauss-Bonnet Best Simplified final form}
\begin{split}
-\frac{1}{2}f\,g_{\mu\nu}+f_R\,R_{\mu\nu}+\big(g_{\mu\nu}\Box-\nabla_\mu\!\nabla_\nu\big)\,f_R
\, =\,\frac{1}{2}\,f_{\!\mathscr{L}_m}\, \big(T_{\mu\nu}-\mathscr{L}_m\, g_{\mu\nu}\big)\;,
\end{split}
\end{equation}
which coincides with the field equation of $\mathscr{L}=f(R,\mathscr{L}_m)$ gravity\cite{f(R Lm)}. Although  a pure Gauss-Bonnet term  in the Lagrangian density cannot change the gravitational field equation $\delta \big(\!\!\sqrt{-g}\,\mathscr{L} \big)/\delta g^{\mu\nu}=0$, it does join the dynamical equation $\delta \big(\!\!\sqrt{-g}\,\mathscr{L} \big)/\delta \phi=0$ when $\mathcal{G}$ is coupled to a scalar field $\phi(x^a)$ (e.g. \cite{GaussBonnet mixed with dynamical scalar field}), and  can still cause nontrivial effects in other aspects (e.g. \cite{Gauss-Bonnet Simplest and Quadratic}).

\subsubsection{{Recovery of} some typical models}

$f(R,\mathcal{G},\mathscr{L}_m)$  is the maximally generalized Gauss-Bonnet gravity when $\{R,\mathcal{G},\mathscr{L}_m\}$
are the only scalar invariants taken into account, and all existing $(R,\mathcal{G},\mathscr{L}_m)$-dependent models can be recovered as a specialized $f(R,\mathcal{G},\mathscr{L}_m)$  gravity. For example,

%\begin{table}
\begin{center}
 \renewcommand\arraystretch{1.69}
\begin{tabular}{|l|l|l|}
  \hline
Reference & Lagrangian density & Specialization \\ \hline\hline
\cite{GaussBonnet first model R/2k+f(G)} & $R/(2\kappa^2)\!+\!f(\mathcal{G})\!+\! \mathscr{L}_m$ & $f_R \!\mapsto\! 1/(2\kappa^2)\;,\:  f_{\mathcal{G}} \!\mapsto\! f_{\mathcal{G}}\;,\:f_{\!\mathscr{L}_m}  \!\mapsto\! 1$  \\  \hline
\cite{Gauss-Bonnet with Nonminimal Coupling f(G)Lm} & $R/2\!+\!\mathscr{L}_m\!+\!\lambda\,f(\mathcal{G})\,\mathscr{L}_m $ &$f_R \!\mapsto\! 1/2\;,\:  f_{\mathcal{G}} \!\mapsto\! \lambda \mathscr{L}_m f_{\mathcal{G}} \;,\:
f_{\!\mathscr{L}_m} \!\mapsto\! 1\!+\!\lambda f(\mathcal{G})$\\   \hline
\cite{Gauss-Bonnet with Nonminimal Coupling f(G)Lm}   & $R/2\!+\!f(\mathcal{G})\!+\!\mathscr{L}_m\!+\!\lambda\,F(\mathcal{G})\,\mathscr{L}_m $ &
$f_R \!\mapsto\! 1/2\;,\:  f_{\mathcal{G}} \!\mapsto\! f_{\mathcal{G}}\!+\!\lambda \mathscr{L}_m F_{\mathcal{G}} \;,\:
f_{\!\mathscr{L}_m} \!\mapsto\! 1\!+\!\lambda F(\mathcal{G})$ \\   \hline
\cite{GaussBonnet second model f(R G)+Lm}   & $f(R, \mathcal{G})\!+\!2\kappa \mathscr{L}_m$ & $f_R \!\mapsto\!  f_R\;,\:  f_{\mathcal{G}} \!\mapsto\! f_{\mathcal{G}}\;,\: f_{\!\mathscr{L}_m}  \!\mapsto\! 2\kappa$\\ \hline
\end{tabular}
\end{center}
%\end{table}

\noindent For a detailed review of generalized Gauss-Bonnet gravity, see \cite{GaussBonnet Review} in which various types of nonminimal coupling are also extensively discussed.

\subsection{Quadratic gravity}

Following the discussion of (generalized) Gauss-Bonnet gravity, we would like to revisit the simplest case with $R_c^2$-dependence (and $R_m^2$-dependence), the so-called quadratic gravity (e.g. \cite{Gauss-Bonnet Simplest and Quadratic}):
\begin{align}
\mathscr{L}\,&=\, R+ \tilde{a}\!\cdot\! R^2 +\tilde{b}\!\cdot\! R_c^2 + \tilde{c}\!\cdot\! R_m^2+\tilde{d}\!\cdot\! R_S^2 +\tilde{e}\!\cdot\!\mathcal {C}^2 +2\kappa \mathscr{L}_m  \label{Quadratic start} \\
&=\, R+ (\tilde{a}-\tilde{c}-\tilde{d}/4-2\tilde{e}/3)\!\cdot\! R^2 +(\tilde{b}+4\tilde{c}+\tilde{d}+2\tilde{e})\!\cdot\! R_c^2 +(\tilde{c}+\tilde{e})\!\cdot\! \mathcal{G} +2\kappa \mathscr{L}_m   \nonumber\\
&\cong\,  R+ a\!\cdot\! R^2 +b\!\cdot\! R_c^2  +2\kappa \mathscr{L}_m   \;. \label{Quadratic Rc2}
\end{align}
The first row is a general linear superposition of some popular quadratic invariants $\{R^2, R_c^2, R_m^2, R_S^2, \mathcal{C}^2\}$ with constant coefficients $\{\tilde{a}, \tilde{b},\ldots\}$, where $\{R_S^2\!=\!R_c^2-R^2/4\; , \mathcal{C}^2\!=\!R_m^2-2R_c^2+R^2/3\}$ respectively denote the square of traceless Ricci tensor and Weyl tensor (see the next subsection).
In Eq.(\ref{Quadratic Rc2}) the pure Gauss-Bonnet term $(\tilde{c}+\tilde{d})\!\cdot\!\mathcal{G}$ has been neglected  for reasons indicated above. Substitution of
\begin{equation}
{f}_R  %\text{ (or }\tilde{f}_R)
\,\mapsto\,1\, ,\quad f_{R^2}\,\mapsto\,a \, ,\quad
f_{R_c^2}\,\mapsto\,b\, ,\quad  f_{R_m^2}\,\mapsto\, 0 \; \;  \mbox{and} \quad f_{\!\mathscr{L}_m}\,\mapsto \,2\kappa
\end{equation}
into Eq.(\ref{FieldEq Explicit R2}) and Eq.(\ref{Hmunu}) yields the field equation  for the quadratic Lagrangian density Eq.(\ref{Quadratic Rc2}),
\begin{equation}\label{Quadratic FieldEqn QRc}
%\begin{split}
-\frac{1}{2}\big(  R+ a\!\cdot\! R^2 +b\!\cdot\! R_c^2 \big)\,g_{\mu\nu}
+\big(1+2aR\big)\,R_{\mu\nu}+2a\,\big(g_{\mu\nu}\Box-\nabla_\mu\!\nabla_\nu\big)\,R+H_{\mu\nu}^{\text{(QRc)}} \,=\,\kappa\,
T_{\mu\nu}\;,
%\end{split}
\end{equation}
where
\begin{equation}
%\begin{split}
H_{\mu\nu}^{\text{(QRc)}}\,=\,  b\!\cdot\!\bigg( 2  R_\mu^{\;\;\,\alpha}R_{\alpha\nu}-  \nabla_\alpha\!\nabla_{\nu} R_{\mu}^{\;\;\,\alpha}-  \nabla_\alpha\!\nabla_{\mu} R_{\nu}^{\;\;\,\alpha}
+   \Box R_{\mu\nu}+  g_{\mu\nu}   \nabla_\alpha\!\nabla_\beta R^{\alpha\beta} \,\bigg)\;.
%\end{split}
\end{equation}
Moreover, via the Bianchi implications Eq.(\ref{Bianchi implications 3}) and Eq.(\ref{Bianchi implications 5}),
$H_{\mu\nu}^{\text{(QRc)}}$ can be rewritten as
\begin{equation}
H_{\mu\nu}^{\text{(QRc)}}\,=\,b\!\cdot\!\bigg( 2 R_{\alpha\mu\beta\nu}R^{\alpha\beta}
+ \big(\frac{1}{2}\,g_{\mu\nu}\Box -\nabla_\mu \nabla_\nu \big)\,R+ \Box R_{\mu\nu}\,\bigg)\;.
\end{equation}
Using this to rewrite Eq.(\ref{Quadratic FieldEqn QRc}), we obtain the commonly used form of the field equation\cite{Gauss-Bonnet Simplest and Quadratic}\cite{Quardratic gravity second paper}.% \\

On the other hand, one can instead drop the Ricci square in favor of the Kretschmann scalar, and accordingly manipulate Eq.(\ref{Quadratic start}) via
\begin{align}
\hspace{10mm}\mathscr{L}&=\,  R+ (\tilde{a}+\tilde{b}/4-\tilde{e}/6)\!\cdot\! R^2 +(\tilde{b}/4+\tilde{c}+\tilde{d}/4+2\tilde{e})/2\!\cdot\! R_m^2 -(\tilde{b}/4+\tilde{d}/4-\tilde{e}/2)\!\cdot\! \mathcal{G} +2\kappa \mathscr{L}_m   \nonumber\\
&\cong\,  R+ a\!\cdot\! R^2 +b\!\cdot\! R_m^2  +2\kappa \mathscr{L}_m   \;.\label{Quadratic Rm2}
\end{align}
Now, substitute ${f}_R \mapsto1$, $f_{R^2}\mapsto a$, $f_{R_c^2}\mapsto 0$, $ f_{R_m^2}\mapsto b$ and  $f_{\!\mathscr{L}_m}\mapsto 2\kappa$ into Eqs.(\ref{FieldEq Explicit R2}) and (\ref{Hmunu}) to obtain
\begin{equation}\label{Quadratic FieldEqn QRm}
%\begin{split}
-\frac{1}{2}\big(R+ a\!\cdot\! R^2 +b\!\cdot\! R_m^2\big)\,g_{\mu\nu}+\big(1+2aR\big)\,R_{\mu\nu}+2b\,\big(g_{\mu\nu}\Box
-\nabla_\mu\!\nabla_\nu\big)\,R +H_{\mu\nu}^{\text{(QRm)}} \,=\,\kappa\,T_{\mu\nu}\;,
%\end{split}
\end{equation}
where
\begin{equation}
%\begin{split}
H_{\mu\nu}^{\text{(QRm)}}\,=\,  b\!\cdot\!\bigg(2  R_{\mu\alpha\beta\gamma}R_{\nu}^{\;\;\,\alpha\beta\gamma}
+4 \nabla^\beta \nabla^\alpha  R_{\alpha\mu \beta\nu} \,\bigg)  \;,\\
%\end{split}
\end{equation}
and $H_{\mu\nu}^{\text{(QRm)}}$ can be recast by the Bianchi property Eq.(\ref{Bianchi implications 5}) into
\begin{equation}
H_{\mu\nu}^{\text{(QRm)}}\,=\,b\!\cdot\!\bigg( 2R_{\mu\alpha\beta\gamma}R_{\nu}^{\;\;\,\alpha\beta\gamma} +4 R_{\alpha\mu\beta\nu}R^{\alpha\beta}
-4 R_{\mu}^{\;\;\,\alpha}R_{\alpha\nu}+4 \Box R_{\mu\nu}-2 \nabla_\mu\! \nabla_\nu R \bigg)\;.
\end{equation}

%Quadratic gravity is the earliest effect in dates back to three decades ago. However, one should be careful that, early works sometimes mistook
%the plus/subtraction sign for some term in $H_{\mu\nu}^{\text{(QRc)}}$ and $H_{\mu\nu}^{\text{(QRm)}}$ in the field equation.

%%%%%%%%%%%%%%%%%%%%%%%%%%%%%%%%%%%%%%%%%%%%%%%%%%%%%%%%%%%%%%%%%%%%%%%%%%%%%%%%%%%%%%%%%%%%
%%%%%%%%%%%%%%%%%%%%%%%%%%%%%%%%%%%%%%%%%%%%%%%%%%%%%%%%%%%%%%%%%%%%%%%%%%%%%%%%%%%%%%%%%%%%
%%%%%%%%%%%%%%%%%%%%%%%%%%%%%%%%%%%%%%%%%%%%%%%%%%%%%%%%%%%%%%%%%%%%%%%%%%%%%%%%%%%%%%%%%%%%

%\subsection{Lanczos-Lovelock Gravity}
%[This small subsection is to be replaced by a remark on Gauss-Bonnet gravity.] The $m=0$ order is a constant. order is a To first order, $m=1$,
%\begin{equation}
%\mathscr{L}_{(LL)}\,=\,\delta^{13}_{24} R^{24}_{13}   \,=\,R_{\mu\nu}-\frac{1}{2}R g_{\mu\nu}
%\end{equation}
%To second order, $m=2$,
%\begin{equation}
%\mathscr{L}_{(LL)}\,=\,\frac{1}{2}\delta^{1357}_{2468} R^{24}_{13} R^{68}_{57}  \,=\,\frac{1}{2} \big( R^2-4R_c^2+R_m^2 \big) \,=\,\frac{1}{2} \,\mathcal G
%\end{equation}
%The first three orders of Lanczos-Lovelock Lagrangian yields the Gauss-Bonnet gravity,
%\begin{equation}
%\mathcal{S}\,=\,\int d^4x \sqrt{-g}\,\big( R_{\mu\nu}-\frac{1}{2}R g_{\mu\nu}+\alpha\,\mathcal G +2\kappa \mathscr{L}_m)  \big)
%\end{equation}

%%%%%%%%%%%%%%%%%%%%%%%%%%%%%%%%%%%%%%%%%%%%%%%%%%%%%%%%%%%%%%%%%%%%%%%%%%%%%%%%%%%%%%%%%%%%
%%%%%%%%%%%%%%%%%%%%%%%%%%%%%%%%%%%%%%%%%%%%%%%%%%%%%%%%%%%%%%%%%%%%%%%%%%%%%%%%%%%%%%%%%%%%

\subsection{Field equations with traceless Ricci and Riemann squares}

It is worthwhile to mention that, as is well known in Riemann geometry, many other tensors can be built algebraically out of $\{R^2,R_{\alpha\beta},R_{\alpha\mu\beta\nu}\}$ with their squares recast into $\{R, R_c^2, R_m^2\}$, such as the traceless Ricci tensor, traceless Riemann tensor (Weyl tensor), Schouten tensor, Plebanski tensor, Bel-Robinson tensor, etc. It can be convenient or sometimes preferable for specific purposes to employ these tensors in the field equation, so in this subsection we will take a quick look at how the squares of these tensors in the Lagrangian density
contribute to the gravitational field equation. It is unnecessary to exhaust all these tensors here and we will just consider the squares of traceless Ricci tensor and Weyl tensor as an example.

\subsubsection{Traceless Ricci square}

The traceless counterpart of Ricci tensor $S_{\alpha\beta}$ ($g^{\alpha\beta}S_{\alpha\beta}=0$) and its square (denoted as $R_S^2$) is,
\begin{equation}
S_{\alpha\beta}\,=\,R_{\alpha\beta}-\frac{1}{4}\,R\,g_{\alpha\beta}\quad\Rightarrow\quad
R_S^2\,\coloneqq\,S_{\alpha\beta}S^{\alpha\beta} = R_c^2-\frac{1}{4}\,R^2 \;.
\end{equation}
Consider $f(\ldots, R_S^2)$ as a generic function of $R_S^2$, where $\ldots$ collects the dependence on all other possible scalar invariants, and the variation $\delta f(\ldots, R_S^2)=\delta f(\ldots, R_c^2-R^2/4)$ yields
\begin{equation}
f_{R_S^2}\!\cdot  \delta R_S^2\,=\,f_{R_S^2}\!\cdot \Big(\frac{\partial R_S^2}{\partial R_c^2}\,\delta R_c^2
+ \frac{\partial R_S^2}{\partial R}\,\delta R  \Big)\,=\,f_{R_S^2}\!\cdot \Big(\delta R_c^2
-\frac{1}{2}\,R\,\delta R  \Big)\;.
\end{equation}
Absorbing $f_{R_S^2}$ into $\delta R_c^2$ by replacing $f_{R_c^2}$ with $f_{R_S^2}$  in Eq.(\ref{Vary Rc2}),
merging $R\,f_{R_S^2}$  into  $\delta R$ by replacing $f_R$ with $R\,f_{R_S^2}$ in Eq.(\ref{Vary f(R)}), and finally replacing all Ricci tensors in $f_{R_S^2} \delta R_c^2$ and $R\,f_{R_S^2}\,\delta R$ by their traceless counterparts $R_{\alpha\beta}\!=\!S_{\alpha\beta}+Rg_{\alpha\beta}/4$,  then $ f_{R_S^2}\!\cdot\!\Big(\delta R_c^2
-\frac{1}{2}\,R\,\delta R  \Big) = f_{R_S^2}\!\cdot\! \delta R_S^2 $ becomes
\begin{equation}\label{Vary RS2 Traceless Ricci square}
\begin{split}
&f_{R_S^2}\!\cdot\! \delta R_S^2
\,=\,\bigg[2f_{R_S^2}\,S_\mu^{\;\;\,\alpha}S_{\alpha\nu}-\frac{1}{2}\,R\,f_{R_S^2}\,S_{\mu\nu}
-\nabla_\alpha\!\nabla_{\nu} \big(S_{\mu}^{\;\;\,\alpha}f_{R_S^2}\big)\\
-&\nabla_\alpha\!\nabla_{\mu} \big(S_{\nu}^{\;\;\,\alpha}f_{R_S^2}\big)
+\Box \big(S_{\mu\nu}f_{R_S^2}\big) +g_{\mu\nu} \nabla_\alpha\!\nabla_\beta \big(S^{\alpha\beta} f_{R_S^2}  \big) \bigg]
\!\cdot\!\delta g^{\mu\nu}\eqqcolon H_{\mu\nu}^{(\!fR_S^2)}\!\cdot\delta g^{\mu\nu}\;,
\end{split}
\end{equation}
which is consistent with the field equation in \cite{Traceless Ricci square model}. Thus, for a Lagrangian density dependent on the traceless Ricci square $\mathscr{L}=f(\ldots, R_S^2)$, the contributions of $f_{R_S^2}\!\cdot\! \delta R_S^2$ to the field equation is just  $H_{\mu\nu}^{(\!fR_S^2)}$ as in Eq.(\ref{Vary RS2 Traceless Ricci square}).

\subsubsection{Weyl square}

Being the totally traceless part of the Riemann tensor in the Ricci decomposition, the Weyl conformal tensor $C_{\alpha\beta\gamma\delta}$ ($g^{\alpha\gamma}g^{\beta\delta}C_{\alpha\beta\gamma\delta}=0$) and its square (denoted as $\mathcal{C}^2$) are respectively
\begin{equation}\label{Weyl tensor and its square I}
C_{\alpha\beta\gamma\delta}\,=\,R_{\alpha\beta\gamma\delta}+\frac{1}{2}\,\Big( g_{\alpha\delta}R_{\beta\gamma}-g_{\alpha\gamma}R_{\beta\delta}+
g_{\beta\gamma}R_{\alpha\delta}- g_{\beta\delta}R_{\alpha\gamma} \Big)+\frac{1}{6}\,\Big( g_{\alpha\gamma} g_{\beta\delta}
- g_{\alpha\delta} g_{\beta\gamma}  \Big)\,R\;,\;\text{ and}
\end{equation}
\begin{flalign}\label{Weyl tensor and its square II}
\mathcal {C}^2\,\coloneqq\, C_{\alpha\mu\beta\nu}C^{\alpha\mu\beta\nu}\,=\,R_m^2-2R_c^2+\frac{1}{3}\,R^2
\,=\,R_m^2-2R_S^2-\frac{1}{6}\,R^2
\,=\,\mathcal{G}+2R_c^2-\frac{2}{3}\,R^2 \;.
\end{flalign}
Given a function $f(\ldots, \mathcal {C}^2 )=f(\ldots, R_m^2-2R_c^2+R^2/3 )=f(\ldots, R_m^2-2R_S^2-R^2/6)=f(\ldots, \mathcal{G}+2R_c^2-2R^2/3)$, the variation $\delta f(\ldots, \mathcal {C}^2 )$ yields
\begin{align}
f_{\mathcal{C}^2}\!\cdot\! \delta C^2\,
\,=\, f_{\mathcal{C}^2}\!\cdot\!\Big(\delta R_m^2-2\,\delta R_c^2
+\frac{2}{3}\,R\,\delta R  \Big)
 \,=\, f_{\mathcal{C}^2}\!\cdot\!\Big(\delta R_m^2-2\,\delta R_S^2
-\frac{1}{3}\,R\,\delta R  \Big)
 \,=\, f_{\mathcal{C}^2}\!\cdot\!\Big(\delta \mathcal{G}+2\,\delta R_c^2-\frac{4}{3}\,R\,\delta R  \Big)\;.
\end{align}
Which of these expressions is  {most convenient to use} will depend on which other Riemann invariants are involved in the Lagrangian density. As such we stop at this stage: the exact expression of $H_{\mu\nu}^{(\!f\mathcal{C}^2)} \delta g^{\mu\nu}\coloneqq f_{\mathcal{C}^2}\!\cdot\! \delta C^2$  depends on which
expansion we choose for $\mathcal{C}^2$.

%Expansion of $\mathcal{C}^2$  should work along well with other Riemannian invariants used in $\mathscr{L}=f(\ldots, \mathcal{C}^2)$, and  $f_{\mathcal{C}^2}\!\cdot\! \delta C^2$ will just contribute to the field equation by the effective tensor $H_{\mu\nu}^{(\!f\mathcal{C}^2)}$.
%
%{\bf ***** }

%%%%%%%%%%%%%%%%%%%%%%%%%%%%%%%%%%%%%%%%%%%%%%%%%%%%%%%%%%%%%%%%%%%%%%%%%%%%%
%%%%%%%%%%%%%%%%%%%%%%%%%%%%%%%%%%%%%%%%%%%%%%%%%%%%%%%%%%%%%%%%%%%%%%%%%%%%%

\section{Nonminimal coupling and energy-momentum divergence}\label{Section 4}

From this section on, we switch our attention to another important aspect of $\mathscr{L}\!=\!f(R,R_c^2,R_m^2,\mathscr{L}_m)$ gravity: the stress-energy-momentum-conservation problem.
Taking the contravariant derivative of the field equation (\ref{FieldEq-2}), we find
\begin{equation}\label{AA generic Multiple f(R Rc2 Rm2 Lm) divergence eqn}
f_{\!\mathscr{L}_m} \nabla^\mu T_{\mu\nu}\,=\,\big(\mathscr{L}_m\, g_{\mu\nu} - T_{\mu\nu}\big)\,\nabla^\mu  f_{\!\mathscr{L}_m}
-  f_{R} \nabla_\nu R -  f_{R_c^2} \nabla_\nu R_c^2
- f_{R_m^2} \nabla_\nu R_m^2  + 2 \nabla^\mu  H_{\mu\nu}^{(\!f R)}+ 2 \nabla^\mu  H_{\mu\nu}^{(\!f R_c^2)}+ 2 \nabla^\mu  H_{\mu\nu}^{(\!f R_m^2)} \;,
\end{equation}
where $\{f,f_R,f_{R_c^2},f_{R_m^2}\}$ remain as functions of the invariants $(R,R_c^2,R_m^2, \mathscr{L}_m)$, and $\{H_{\mu\nu}^{(\!fR)}, H_{\mu\nu}^{(\!fR_c^2)},H_{\mu\nu}^{(\!fR_m^2)}\}$ have already been concretized in Eqs.(\ref{Vary f(R)})-(\ref{Vary Rm2}). However, despite the extended variable-dependence in $f_R(R,R_c^2,R_m^2, \mathscr{L}_m)$ as opposed to $f(R)+2\kappa\mathscr{L}_m$ gravity, we still have\footnote{This is actually the stress-energy-momentum conservation condition of $f(R)$ gravity with Lagrangian density
$\mathscr{L}=f(R)+2\kappa \mathscr{L}_m$ and field equation  $-f(R)\,g_{\mu\nu}/2+f_R R_{\mu\nu}+(g_{\mu\nu}\Box-\nabla_\mu\nabla_\nu)f_R=\kappa T_{\mu\nu}$ , except that $f_R=f_R(R)$. }
\begin{equation}\label{AA generic f(R) conserve}
\frac{1}{2}\,\bigg(\!-  f_{R} \nabla_\nu R+ 2 \nabla^\mu  H_{\mu\nu}^{(\!f R)}\bigg)\,=\,
-f_R\, \nabla^\mu \Big(\frac{1}{2}\,R\,g_{\mu\nu}\Big)+\nabla^\mu \big(\,f_R\!\cdot\!R_{\mu\nu}\, \big)+\big(\nabla_\nu \Box-\Box\nabla_\nu\big)\,f_R \,=\,0\;.
\end{equation}
It vanishes as a consequence of the contracted Bianchi identity $\nabla^\mu(R_{\mu\nu}-R g_{\mu\nu}/2)=0$ and the third-order-derivative commutation relation $(\Box\nabla_\nu -\nabla_\nu \Box)f_R=R_{\mu\nu}\nabla^\nu f_R$ .
Thus, Eq.(\ref{AA generic Multiple f(R Rc2 Rm2 Lm) divergence eqn}) further reduces to
\begin{equation}\label{Divergence of StressEnergyMomentum}
f_{\!\mathscr{L}_m}\nabla^\mu T_{\mu\nu}\,=\,\big(\mathscr{L}_m\, g_{\mu\nu} - T_{\mu\nu}\big)\,\nabla^\mu f_{\!\mathscr{L}_m}
-  f_{R_c^2} \nabla_\nu R_c^2
- f_{R_m^2} \nabla_\nu R_m^2  + 2 \nabla^\mu  H_{\mu\nu}^{(\!f R_c^2)}+ 2 \nabla^\mu  H_{\mu\nu}^{(\!f R_m^2)} \;,
\end{equation}
which constitutes the equation of energy-momentum divergence   in $f(R,R_c^2,R_m^2,\mathscr{L}_m)$ gravity. It can be regarded as a generalization of the following divergence equation in $f(R,\mathscr{L}_m)$ gravity\cite{f(R Lm)},
\begin{equation}\label{AA generic f(R Lm) divergence eqn}
\nabla^\mu T_{\mu\nu}\,=\,  \big(\mathscr{L}_m\, g_{\mu\nu}-T_{\mu\nu} \big)\,\nabla^\mu \ln f_{\!\mathscr{L}_m}\;,
\end{equation}
with $\nabla^\mu \ln f_{\!\mathscr{L}_m}\equiv f_{\!\mathscr{L}_m}^{-1}\nabla^\mu  f_{\!\mathscr{L}_m}$, which in turn can be recovered from Eq.(\ref{Divergence of StressEnergyMomentum}) by setting $f_{R_c^2}=0=f_{R_m^2}$.% \\

In standard GR, $\nabla^\mu T_{\mu\nu}=0$ is the mathematical expression of conservation of stress-energy-momentum. However for our models it is clear that this does not vanish and so this fundamental conservation law does not hold in the standard form. Then, how to understand the energy-momentum nonconservation/divergence equation (\ref{Divergence of StressEnergyMomentum})? Is it further reducible and how does it influence the equations of continuity and motion given concrete matter sources? We will investigate these questions in a more generic framework.

\subsection{Automatic energy-momentum conservation under minimal coupling}\label{AA automati conservation minimal coupling}

Consider a generic gravitational Lagrangian $\mathscr{L}_G=f(\mathcal{R})$ where  $f(\mathcal{R})$ is an arbitrary function of an $(n+2)$-order algebraic $(n=0)$ or differential ($n\geq 1$) Riemannian invariant $\mathcal{R}$:
\begin{equation}\label{AA generic Riemannian invariant R}
\mathcal{R}\,=\,\mathcal{R}(g_{\alpha\beta}\,,R_{\alpha\mu\beta\nu}\,,\nabla_\gamma R_{\alpha\mu\beta\nu}\,,
\ldots\,,\nabla_{\gamma_ 1}\!\nabla_{\gamma_ 2}\ldots \nabla_{\gamma_ n} R_{\alpha\mu\beta\nu})\;,
\end{equation}
so that variational derivative of the covariant density $\sqrt{-g}\,\mathscr{L}_G$ will lead to a $(2n+4)$-order model of gravity. Such an $\mathscr{L}_G=f(\mathcal{R})$ is still a covariant invariant for which Noether's conservation law would yield\cite{Generalized Bianchi and Conservation 1}
\begin{equation}\label{AA generalized bianchi identities}
\nabla^\mu \,\left( \frac{1}{\sqrt{-g}} \,\frac{\delta\, \Big(\!\!\sqrt{-g}\,f(\mathcal{R}) \Big)}{\delta g^{\mu\nu}}\right)=\,0\;,
\end{equation}
which can be expanded into
\begin{equation}\label{AA generalized bianchi identities Equivalent}
 f_{\mathcal{R}}(\mathcal{R})\cdot \! \nabla_\nu \mathcal{R}
\,=\, 2\nabla^\mu  H_{\mu\nu}^{(\!f{\mathcal{R}})}\quad\text{ with } \quad
H_{\mu\nu}^{(\!f{\mathcal{R}})}\!\cdot\delta g^{\mu\nu}\,\coloneqq\,f_{\mathcal{R}}\!\cdot\delta \mathcal{R}\;,
\end{equation}
where $H_{\mu\nu}^{(\!f{\mathcal{R}})}$  is defined the same way as $\{H_{\mu\nu}^{(\!fR)}, H_{\mu\nu}^{(\!fR_c^2)},H_{\mu\nu}^{(\!fR_m^2)}\}$ in
Eqs.(\ref{Vary f(R)})-(\ref{Vary Rm2}). It absorbs $f_{\mathcal{R}}$ into $\delta \mathcal{R}$ and collects all nonlinear and higher-order terms
generated by $f_{\mathcal{R}}\!\cdot\delta \mathcal{R}$.% \\

These results can be directly generalized to the situation where  $\mathscr{L}_G$ relies on multiple Riemannian invariants,
$\mathscr{L}_G=f(\mathcal{R}_1,\mathcal{R}_2,\ldots,\mathcal{R}_p)\equiv\mathscr{L}_G(g_{\alpha\beta},R_{\alpha\mu\beta\nu},\nabla_\gamma R_{\alpha\mu\beta\nu},\ldots,\nabla_{\gamma_ 1}\!\nabla_{\gamma_ 2}\ldots \nabla_{\gamma_q} R_{\alpha\mu\beta\nu})$, and we have
\begin{equation}\label{AA generalized bianchi identities Equivalent II multiple}
\sum_i f_{\mathcal{R}_i}\,  \nabla_\nu \mathcal{R}_i
\,=\, 2\,\sum_i \nabla^\mu  H_{\mu\nu}^{(\!f{\mathcal{R}}_i)}\quad\text{ with }\quad
H_{\mu\nu}^{(\!f{\mathcal{R}_i})}\!\cdot \delta g^{\mu\nu}\,\coloneqq\,f_{\mathcal{R}_i}\!\cdot  \delta \mathcal{R}_i \;,
\end{equation}
where $f_{\mathcal{R}_i}=f_{\mathcal{R}_i}(\mathcal{R}_1\,,\mathcal{R}_2\,,\ldots\,,\mathcal{R}_p)$, with each $\mathcal{R}_i$ given by Eq.(\ref{AA generic Riemannian invariant R}) to certain order derivatives of Riemann tensor, and $H_{\mu\nu}^{(\!f{\mathcal{R}}_i)}=H_{\mu\nu}^{(\!f{\mathcal{R}}_i)}(\mathcal{R}_1\,,\mathcal{R}_2\,,\ldots\,,\mathcal{R}_p)$ absorbs $f_{\mathcal{R}_i}$ into $ \delta \mathcal{R}_i$.% \\

Since $f(\mathcal{R}_1,\mathcal{R}_2,\ldots,\mathcal{R}_p)$ is a purely  geometric entity solely dependent on the metric and derivatives of Riemann tensor,  Eqs.(\ref{AA generalized bianchi identities Equivalent}) and (\ref{AA generalized bianchi identities Equivalent II multiple})
arising from Noether's theorem are also called the ``generalized (contracted) Bianchi identities''\cite{Generalized Bianchi and Conservation 1}\cite{Generalized Bianchi and Conservation 3}. As the simplest example, when $f(\mathcal{R}_1,\mathcal{R}_2,\ldots,\mathcal{R}_p)=R$,  Eq.(\ref{AA generalized bianchi identities Equivalent}) or Eq.(\ref{AA generalized bianchi identities Equivalent II multiple}) immediately reproduces the standard contracted Bianchi identity $\nabla^\mu(R_{\mu\nu}-Rg_{\mu\nu}/2)=0$ which is often used in GR. % \\

On the other hand, for the matter Lagrangian density $\mathscr{L}_m$, Noether's conservation law yields
\begin{equation}\label{AA minimally coupled Tmunu}
\nabla^\mu \,\left( \frac{1}{\sqrt{-g}} \,\frac{\delta\, \Big(\!\!\sqrt{-g}\,\mathscr{L}_m \Big)}{\delta g^{\mu\nu}}\right)=\,0\,=\,-\frac{1}{2}\nabla^\mu T_{\mu\nu}\quad\text{ with }\quad
T_{\mu\nu}\coloneqq \frac{-2}{\sqrt{-g}} \,\frac{\delta\, \Big(\!\!\sqrt{-g}\,\mathscr{L}_m \Big)}{\delta g^{\mu\nu}}\;,
\end{equation}
where $T_{\mu\nu}$ is the standard stress-energy-momentum (SEM) tensor as in Eq.(\ref{Tmunu Definition}). This way of defining $T_{\mu\nu}$ from Noether's law therefore naturally guarantees  energy-momentum conservation $\nabla^\mu T_{\mu\nu}=0$. Moreover, in the case of minimal  coupling, it is unnecessary to consider a covariant matter density of the form $\sqrt{-g}\,h(\mathscr{L}_m)$, since $h(\mathscr{L}_m)$ can always be treated as a whole, $h(\mathscr{L}_m)\mapsto \tilde{\mathscr{L}}_m$.% \\

Hence, for a generic Lagrangian density where $\mathscr{L}_m$ is minimally coupled to the spacetime geometry:
\begin{equation}\label{AA generic minimal coupled Lagrangian density}
\mathscr{L}=\mathscr{L}_G+2\kappa \mathscr{L}_m= f(\mathcal{R}_1\,,\mathcal{R}_2\,,\ldots\,,\mathcal{R}_p)+2\kappa \mathscr{L}_m\;,
\end{equation}
and whose field equation arises from  extremizing the action or equivalently $\displaystyle \frac{1}{\sqrt{-g}}\frac{\delta\, (\!\sqrt{-g}\,\mathscr{L})}{\delta g^{\mu\nu}}=0$:
\begin{equation}\label{AA generic minimal coupled field eqn}
-\frac{1}{2}fg_{\mu\nu}+\sum_i H_{\mu\nu}^{(\!f{\mathcal{R}}_i)}  \,=\,\kappa T_{\mu\nu}\;,
\end{equation}
the generalized Bianchi identities Eq.(\ref{AA generalized bianchi identities Equivalent II multiple}) for pure geometric $\mathscr{L}_G$ together with the Noether-type definition of $T_{\mu\nu}$ in Eq.(\ref{AA minimally coupled Tmunu}) yield that contravariant derivatives of  the left (geometry) and right (matter) -hand side of the field equation (\ref{AA generic minimal coupled field eqn}) vanish  \emph{independently}\footnote{{Instead of  directly starting from Eq.(\ref{Tmunu Definition}), one can consider $T_{\mu\nu}$ from the perspective of diffeomorphism (or gauge) invariance by requiring that the total action $\mathcal{S}_G+\mathcal{S}_m$ be invariant under an arbitrary and  infinitesimal active transformation $g_{\mu\nu}\mapsto g_{\mu\nu}+\delta_\zeta g_{\mu\nu}= g_{\mu\nu}+\nabla_\mu \zeta_\nu+\nabla_\nu \zeta_\mu$, where $\zeta^\mu$ vanishes at the boundary.
\begin{equation}
\delta\mathcal{S}_m=-\frac{1}{2}\delta\!\int d^4x \sqrt{-g}\,T_{\mu\nu}\,\delta g^{\mu\nu}=-\delta\!\int d^4x \sqrt{-g}\,T_{\mu\nu}\,\nabla^\mu\zeta^\nu\,\cong\,\delta\!\int d^4x \sqrt{-g}\,(\nabla^\mu T_{\mu\nu})\,\zeta^\nu\;.
\end{equation}
Now the automatic conservation $\nabla^\mu T_{\mu\nu}=0$ would become a consequence of the (generalized) Bianchi identities which arise from the diffeomorphism invariance of $\mathcal{S}_G$. Both ways trace back to Noether's law.}}. This ensures automatic fulfillment of  energy-momentum conservation in any minimally coupled gravity theories of the form Eqs.(\ref{AA generic minimal coupled Lagrangian density}) and (\ref{AA generic minimal coupled field eqn}), such as $\mathscr{L}\!=\!f(R,R_c^2,R_m^2)\!+\!2\kappa\mathscr{L}_m$ gravity and $\mathscr{L}\!=\!f(R,\mathcal{G})\!+\!2\kappa\mathscr{L}_m$ gravity.

%%%%%%%%%%%%%%%%%%%%%%%%%%%%%%%%%%%%%%%%%%%%%

%%%%%%%%%%%%%%%%%%%%%%%%%%%%%%%%%%%%%%%%%%%%%

\subsection{Divergence of SEM tensor under nonminimal coupling}\label{AA nonminimal coiupling Section}

Now consider a generic  Lagrangian density $\mathscr{L}=f(\mathcal{R}_1,\ldots,\mathcal{R}_p,\mathscr{L}_m)$ which allows nonminimal coupling between $\mathscr{L}_m$ and Riemannian  invariants $\mathcal{R}_i$.  Noether's law yields the following equation for the divergence of the energy-momentum tensor,
\begin{equation}%\label{AA generic Multiple nonminimal coupling divergence eqn}
\nabla^\mu \,\left( \frac{1}{\sqrt{-g}} \,\frac{\delta\, \Big(\!\!\sqrt{-g}\,f(\mathcal{R}_1, \ldots,\mathcal{R}_p,\mathscr{L}_m) \Big)}{\delta g^{\mu\nu}}\right)=0\;,
\end{equation}
with expansion
\begin{equation}\label{AA generic Multiple nonminimal coupling divergence eqn}
f_{\!\mathscr{L}_m} \nabla^\mu T_{\mu\nu}=\big(\mathscr{L}_m\, g_{\mu\nu} - T_{\mu\nu}\big)\,\nabla^\mu  f_{\!\mathscr{L}_m}
- \sum_i  f_{\mathcal{R}_i}  \nabla_\nu \mathcal{R}_i  + 2\sum_i \nabla^\mu  H_{\mu\nu}^{(\!f{\mathcal{R}}_i)}\;,
\end{equation}
where \{$f_{\!\mathscr{L}_m}$, $f_{\mathcal{R}_i}$\} are all dependent on $(\mathcal{R}_1\ldots,\mathcal{R}_p,\mathscr{L}_m)$, and $H_{\mu\nu}^{(f{\mathcal{R}_i})} \delta g^{\mu\nu}\!\coloneqq\! f_{\mathcal{R}_i}\delta \mathcal{R}_i$ as usual. Note that, ``conservation'' of $\sqrt{-g}\,f(, \ldots,\mathcal{R}_p,\mathscr{L}_m)$ yields an
 unavoidable ``divergence'' term $\big(\mathscr{L}_m g_{\mu\nu} \!-\! T_{\mu\nu}\big)\nabla^\mu  f_{\!\mathscr{L}_m}$ essentially because of how $T_{\mu\nu}$ was defined; that is to say, for the nonminimally coupled $\mathscr{L}\!=\!f(\mathcal{R}_1,\ldots,\mathcal{R}_p,\mathscr{L}_m)$ under discussion, we have continued to use the definition of $T_{\mu\nu}$ from Eq.(\ref{AA minimally coupled Tmunu}) which was adapted to minimal coupling. Also, for $\mathscr{L}=f(R, \mathcal{R}_1,\ldots,\mathcal{R}_p,\mathscr{L}_m)$ gravity where the first invariant is identified as the Ricci scalar, the same argument as Eq.(\ref{AA generic f(R) conserve}) yields that $-f_R \nabla_\nu R+H_{\mu\nu}^{(fR)}=0$ for $f_R\!=\!f_R(R, \mathcal{R}_1,\ldots,\mathcal{R}_p,\mathscr{L}_m)$.% \\

For the moment, we cannot directly use Eq.(\ref{AA generalized bianchi identities Equivalent II multiple}) to eliminate $- \sum_i  f_{\mathcal{R}_i}  \nabla_\nu \mathcal{R}_i$ by $2\sum_i \nabla^\mu  H_{\mu\nu}^{(\!f{\mathcal{R}}_i)}$ in Eq.(\ref{AA generic Multiple nonminimal coupling divergence eqn}) as they are no longer purely geometric entities. In principle, the coefficient $f_{\mathcal{R}_i}=f_{\mathcal{R}_i}(\mathcal{R}_1,\ldots,\mathcal{R}_p,\mathscr{L}_m)$ allows for arbitrary dependence on $\mathscr{L}_m$, and this complexity gets even further promoted after taking the contravariant derivative of the effective tensor $ H_{\mu\nu}^{(\!f{\mathcal{R}}_i)}(f_{\mathcal{R}_i})$. Also, note that, for the Lagrangian density $\mathscr{L}=f(R,R_c^2,R_m^2,\mathscr{L}_m)$ and $\mathscr{L}=f(R,\mathscr{L}_m)$ , the generic result Eq.(\ref{AA generic Multiple nonminimal coupling divergence eqn}) soon recovers Eqs.(\ref{AA generic Multiple f(R Rc2 Rm2 Lm) divergence eqn}) and (\ref{AA generic f(R Lm) divergence eqn}), which were obtained in an alternative way from directly taking contravariant derivatives of their field equation.% \\

 As we have already learned, in Eq.(\ref{AA generic Multiple nonminimal coupling divergence eqn}) the  term $\big(\mathscr{L}_m g_{\mu\nu} - T_{\mu\nu}\big)\nabla^\mu  f_{\!\mathscr{L}_m}$ originates from the contradiction between the nonminimal $\mathcal{R}_i-\mathscr{L}_m$ coupling and the minimal definition of $T_{\mu\nu}$. However, how can we understand the other divergence terms $- \sum_i  f_{\mathcal{R}_i}  \nabla_\nu \mathcal{R}_i $ and $2\sum_i \nabla^\mu  H_{\mu\nu}^{(\!f{\mathcal{R}}_i)}$? Fortunately, investigations of $\mathscr{L}\!=\!\tilde{f}(\mathcal{R})\!+\!2\kappa \mathscr{L}_m\!+\!f(\mathcal{R})\,\mathscr{L}_m$ gravity shed some light  on this question.

%%%%%%%%%%%%%%%%%%%%%%%%%%%%%%%%%%%%%%%%%%%%%

%%%%%%%%%%%%%%%%%%%%%%%%%%%%%%%%%%%%%%%%%%%%%

\subsection{Lessons from $\tilde{f}(\mathcal{R}_i)\!+\!2\kappa \mathscr{L}_m \!+\!f(\mathcal{R}_i)\mathscr{L}_m$ model}

Now, consider a further specialized model with Lagrangian density
\begin{equation}
\mathscr{L}=\tilde{f}(\mathcal{R}_1,\ldots,\mathcal{R}_p)+2\kappa \mathscr{L}_m+f(\mathcal{R}_1,\ldots,\mathcal{R}_q)\cdot\!\mathscr{L}_m\;.
\end{equation}
Sec.~\ref{AA automati conservation minimal coupling} has shown us that,  energy-momentum conservation (divergence-freeness) is automatically satisfied for the minimally coupled component $\tilde{f}(\mathcal{R}_1,\ldots,\mathcal{R}_p)+2\kappa \mathscr{L}_m$, so we just need to concentrate on the nonminimally coupled term $f(\mathcal{R}_1,\ldots,\mathcal{R}_q)\cdot\!\mathscr{L}_m$. Following the discussion in Sec.~\ref{AA nonminimal coiupling Section} just above,
treat $f(\mathcal{R}_1,\ldots,\mathcal{R}_q)\cdot\!\mathscr{L}_m$ as an invariant, so that Noether conservation of the covariant Lagrangian density
$\sqrt{-g}f(\mathcal{R}_1,\ldots,\mathcal{R}_q)\cdot\!\mathscr{L}_m$ yields
\begin{equation}
\nabla^\mu \,\left( \frac{1}{\sqrt{-g}} \,\frac{\delta\, \Big(\!\!\sqrt{-g}\,f(\mathcal{R}_1,\ldots,\mathcal{R}_q)\cdot\!\mathscr{L}_m \Big)}{\delta g^{\mu\nu}}\right)=\,0\;,
\end{equation}
which in turn implies that
\begin{equation}\label{AA Example generic nonminimal coupling divergence eqn}
f \, \nabla^\mu T_{\mu\nu}\,=\,\big(\mathscr{L}_m\, g_{\mu\nu} - T_{\mu\nu}\big)\,\nabla^\mu  f
- \sum_i  f_{\mathcal{R}_i}(\mathcal{R}_1,\ldots,\mathcal{R}_q)\cdot\!  \nabla_\nu \mathcal{R}_i
+ 2\sum_i \nabla^\mu  \left(\frac{\mathscr{L}_m f_{\mathcal{R}_i}\!\cdot  \delta \mathcal{R}_i }{\delta g^{\mu\nu}} \right)\;.
\end{equation}
Note that in the last term, $\mathscr{L}_m f_{\mathcal{R}_i}(\mathcal{R}_1,\ldots,\mathcal{R}_q)\!\cdot\!  \delta \mathcal{R}_i $ acts as a unity rather than a triple multiplication and \emph{cannot} be expanded via the product rule when acted upon by $\nabla^\mu$: In fact, $\mathscr{L}_m f_{\mathcal{R}_i}(\mathcal{R}_1,\ldots,\mathcal{R}_q)\!\cdot\!  \delta \mathcal{R}_i \eqqcolon H_{\mu\nu}^{(\mathscr{L}_mf{\mathcal{R}}_i)} \!\cdot   \delta g^{\mu\nu}$ and thus $\mathscr{L}_m f_{\mathcal{R}_i}$ is merged into $\delta \mathcal{R}_i$. % \\

Now recall that, based on the Petrov and Serge classifications, there are fourteen independent algebraic Riemannian invariants $\mathcal{I}=\mathcal{I}\big(g_{\alpha\beta}, R_{\alpha\mu\beta\nu}\big)$ characterizing a four-dimensional spacetime\cite{Riemann invariants}\cite{Riemann invariants II},  among which nine are of even parity and five are of odd parity, though this minimum set can be slightly expanded after considering the matter content. As a special example of Eq.(\ref{AA Example generic nonminimal coupling divergence eqn}), energy-momentum  divergence of the nonminimally coupled Lagrangian $f(\mathcal{I}_1,\ldots,\mathcal{I}_9)\cdot\mathscr{L}_m$ was studied  in \cite{Nonminimal Coupling I}, where $\{\mathcal{I}_1,\ldots,\mathcal{I}_9\}$ refer to the nine parity-even algebraic Riemannian invariants. Explicit calculations of $H_{\mu\nu}^{(\mathscr{L}_mf{\mathcal{I}_i})}$ and $\nabla^\mu H_{\mu\nu}^{(\mathscr{L}_mf{\mathcal{I}_i})}$ show that\cite{Nonminimal Coupling I}, for each individual $\mathcal{I}_i$ in $\mathscr{L}\!=\!f(\mathcal{I}_i\,,\mathscr{L}_m)$,
\begin{equation}
-   f_{\mathcal{I}_i}(\mathcal{I}_i)\cdot\!  \nabla_\nu \mathcal{I}_i
+ 2\nabla^\mu  \left(\frac{\mathscr{L}_m f_{\mathcal{I}_i}(\mathcal{I}_i)\!\cdot\!  \delta \mathcal{I}_i }{\delta g^{\mu\nu}} \right)\,=\,0\;,
\end{equation}
and most generally for $f(\mathcal{I}_1,\ldots,\mathcal{I}_9)\cdot\!\mathscr{L}_m$ with an arbitrary multiple dependence of these nine invariants,
\begin{equation}
-   \sum_i f_{\mathcal{I}_i}(\mathcal{I}_1,\ldots,\mathcal{I}_9)\cdot\!  \nabla_\nu \mathcal{I}_i
+ 2\sum_i \nabla^\mu  \left(\frac{\mathscr{L}_m f_{\mathcal{I}_i}(\mathcal{I}_1,\ldots,\mathcal{I}_9)\!\cdot\!  \delta \mathcal{I}_i }{\delta g^{\mu\nu}} \right)\,=\,0\;.
\end{equation}
Hence,  the equation of energy-momentum divergence for $\mathscr{L}\!=\!\tilde{f}(\mathcal{I}_1,\ldots,\mathcal{I}_9)+2\kappa \mathscr{L}_m+f(\mathcal{I}_1,\ldots,\mathcal{I}_9) \cdot \mathscr{L}_m$ gravity finally becomes
\begin{equation}
f(\mathcal{I}_1,\ldots,\mathcal{I}_9) \cdot\! \nabla^\mu T_{\mu\nu}\,=\,\big(\mathscr{L}_m\, g_{\mu\nu} - T_{\mu\nu}\big) \cdot\! \nabla^\mu  f(\mathcal{I}_1,\ldots,\mathcal{I}_9)\;.
\end{equation}

\subsection{Conjecture for energy-momentum divergence}

Now, let's summarize the facts we have confirmed so far:

\begin{enumerate}
  \item  In  the simplest $\mathscr{L}\!=\!f(R,\mathscr{L}_m)$ gravity\cite{f(R Lm)}, one has $-  f_{R} \nabla_\nu R+ 2 \nabla^\mu  H_{\mu\nu}^{(\!f R)}=0$, so  $R$-dependence in $\mathscr{L}\!=\!f$  makes no contribution and $\big(\mathscr{L}_m\, g_{\mu\nu} - T_{\mu\nu}\big)\,\nabla^\mu f_{\!\mathscr{L}_m}$ is the only energy-momentum divergence term;
  \item In $\mathscr{L}\!=\!f(R,\mathcal{R}_1,\mathcal{R}_2,\ldots,\mathcal{R}_p,\mathscr{L}_m)$ gravity,  $-  f_{R} \nabla_\nu R+ 2 \nabla^\mu  H_{\mu\nu}^{(\!f R)}=0$ for $f_{R}=f_{R}(R,\mathcal{R}_1,\mathcal{R}_2,\ldots,\mathcal{R}_p,\mathscr{L}_m)$;
  \item In $\mathscr{L}\!=\!\tilde{f}(\mathcal{I}_1,\ldots,\mathcal{I}_9)+2\kappa \mathscr{L}_m+f(\mathcal{I}_1,\ldots,\mathcal{I}_9)\cdot\!\mathscr{L}_m$ gravity\cite{Nonminimal Coupling I}, one has individually $- f_{\mathcal{I}_i}(\mathcal{I}_i)\cdot\!  \nabla_\nu \mathcal{I}_i
+ 2\nabla^\mu H_{\mu\nu}^{(\mathscr{L}_mf{\mathcal{I}i})}=0$ and collectively $-\sum_i f_{\mathcal{I}}(\mathcal{I}_i)\cdot\!  \nabla_\nu \mathcal{I}_i
+ 2\sum_i \nabla^\mu H_{\mu\nu}^{(\mathscr{L}_mf{\mathcal{I}i})}=0$, so $\big(\mathscr{L}_m\, g_{\mu\nu} - T_{\mu\nu}\big)\,\nabla^\mu f_{\!\mathscr{L}_m}$ is the only nonconservation term, while $\mathcal{I}_i$-dependence in $f\!\cdot\!\mathscr{L}_m$ makes no contribution;
  \item In the case of minimal coupling, all algebraic and differential Riemannian invariants $\mathcal{R}_i$ act equally and indiscriminately  in front of Noether's conservation law and generalized Bianchi identities.
\end{enumerate}

\noindent  Starting with these results, the belief that for the situation of generic nonminimal curvature-matter coupling all Riemannian invariants continue to play equal roles in energy-momentum nonconservation/divergence leads us to propose the following:  \\

\noindent  \emph{Weak conjecture}: Consider a Lagrangian density allowing generic nonminimal coupling between the matter density $\mathscr{L}_m$ and Riemannian invariants $\mathcal{R}$,
\begin{equation}\label{AA conjecture nionminimal coupling Lagrangian}
\mathscr{L}\,=\,f(\mathcal{R}_1\,,\mathcal{R}_2\ldots,\mathcal{R}_n\,,\mathscr{L}_m)\;,
\end{equation}
where
$$
\mathcal{R}_i\,=\,\mathcal{R}_i\,\big(g_{\alpha\beta}\,,R_{\alpha\mu\beta\nu}\,,\nabla_\gamma R_{\alpha\mu\beta\nu}\,,
\ldots\,,\nabla_{\gamma_ 1}\!\nabla_{\gamma_ 2}\ldots \nabla_{\gamma_ m} R_{\alpha\mu\beta\nu}\big)\;.
$$
Then contributions from the $\mathcal{R}_i$-dependence of $\mathscr{L}\!=\!f$ in the Noether-induced divergence equation cancel out collectively,
\begin{equation}\label{AA Tian-Booth weak conjecture condition}
-   \sum_i f_{\mathcal{R}_i}\! \cdot \nabla_\nu \mathcal{R}_i
+ 2\sum_i \nabla^\mu   H_{\mu\nu}^{(\!f\mathcal{R}_i)}  \,=\,0\;,
\end{equation}
and the equation of energy-momentum conservation/divergence takes the form\footnote{When talking about its nontrivial divergence, $T_{\mu\nu}$ can be  understood as the $T_{\mu\nu}^{\text{(NC)}}$ which comes from the $\mathscr{L}_m$ under nonminimal coupling, because the contribution $T_{\mu\nu}^{\text{(MC)}}$ to the total SEM tensor by an isolated (i.e. minimally coupled) covariant matter density $\sqrt{-g}\,\mathscr{L}_m$ automatically satisfies the standard stress-energy-momentum conservation. }
\begin{equation}\label{AA conjecture divergence eqn}
f_{\!\mathscr{L}_m}\! \cdot\nabla^\mu T_{\mu\nu}\,=\,\big(\mathscr{L}_m\, g_{\mu\nu} - T_{\mu\nu}\big)\,\nabla^\mu f_{\!\mathscr{L}_m}\;,
\end{equation}
where $\displaystyle H_{\mu\nu}^{(\!f\mathcal{R}_i)}\coloneqq \frac{f_{\mathcal{R}_i}(\mathcal{R}_1,\ldots,\mathscr{L}_m)\!\cdot\!  \delta \mathcal{R}_i }{\delta g^{\mu\nu}}$ ,
$f_{\mathcal{R}_i}=f_{\mathcal{R}_i}(\mathcal{R}_1,\ldots,\mathscr{L}_m)$ ,  and $f_{\!\mathscr{L}_m}=f_{\!\mathscr{L}_m}(\mathcal{R}_1,\ldots,\mathcal{R}_n,\mathscr{L}_m)$.% \\

Moreover, inspired by the behavior of $R$ in Eq.(\ref{AA generic f(R) conserve}) that $-f_{R} \nabla_\nu R \!+\! 2 \nabla^\mu H_{\mu\nu}^{(\!f R)}=0$ in spite of $f_R=f_R(R,R_c^2,R_m^2, \mathscr{L}_m)$, we further promote the weak conjecture to the following: \\

\noindent  \emph{Strong conjecture}: For every invariant $\mathcal{R}_i$ in $\mathscr{L}\!=\!f(\mathcal{R}_1,\mathcal{R}_2\ldots,\mathcal{R}_n,\mathscr{L}_m)$, the divergence terms arising from each $\mathcal{R}_i$-dependence in $\mathscr{L}\!=\!f$ cancel out \emph{individually},
\begin{equation}\label{AA Tian-Booth strong conjecture condition}
-   f_{\mathcal{R}_i}\! \cdot \nabla_\nu \mathcal{R}_i
+ 2\nabla^\mu H_{\mu\nu}^{(\!f\mathcal{R}_i)} \,=\,0\;,
\end{equation}
and the equation of energy-momentum conservation/divergence remains the same as in Eq.(\ref{AA conjecture divergence eqn}),
$$
f_{\!\mathscr{L}_m}\! \cdot\nabla^\mu T_{\mu\nu}\,=\,\big(\mathscr{L}_m\, g_{\mu\nu} - T_{\mu\nu}\big)\,\nabla^\mu f_{\!\mathscr{L}_m}\;.
$$

\vspace{2mm}Specifically, when the possible nonminimal coupling reduces to ordinary  minimal coupling, Eq.(\ref{AA conjecture nionminimal coupling Lagrangian}) will be specialized into $\mathscr{L}\!=\! f(\mathcal{R}_1,\ldots,\mathcal{R}_n)\!+\!2\kappa \mathscr{L}_m$ as in Eq.(\ref{AA generic minimal coupled Lagrangian density}), so  Eqs.(\ref{AA Tian-Booth weak conjecture condition}) and (\ref{AA Tian-Booth strong conjecture condition}) in the weak conjecture are naturally satisfied because of the generalized Bianchi identities Eqs.(\ref{AA generalized bianchi identities Equivalent}) and (\ref{AA generalized bianchi identities Equivalent II multiple}). Also,  if the conjecture were correct, then the generalized Bianchi indentities Eqs.(\ref{AA generalized bianchi identities Equivalent}) and (\ref{AA generalized bianchi identities Equivalent II multiple}) could be generalized again, and they cannot serve as a sufficient condition for judging minimal coupling. % \\

Furthermore,
%\begin{equation}
%\frac{f_{\!\mathscr{L}_m}}{\mathscr{L}_m\, g_{\mu\nu} - T_{\mu\nu}}  \nabla^\mu T_{\mu\nu}\,=\,\nabla^\mu f_{\!\mathscr{L}_m}\;,
%\end{equation}
{reading left to right the nonconservation equation (\ref{AA conjecture divergence eqn})}  clearly shows that the energy-momentum divergence is transformed into the gradient of nonminimal gravitational coupling strength $f_{\!\mathscr{L}_m}$.
On the other hand, if the weak or even the strong conjecture were true,  does it mean that differences between the set of Riemannian invariants which the Lagrangian density depends on are trivial? The answer is of course no, because the gradient $ \nabla^\mu f_{\!\mathscr{L}_m}$ is superposed by the gradient of  $\mathscr{L}_m$ and the gradients of all characteristic Riemannian invariants $\mathcal{R}_i$ used in $\mathscr{L}=f$:
\begin{equation}\label{AA second minimal coupling condition}
f_{\!\mathscr{L}_m}\! \cdot \nabla^\mu T_{\mu\nu}\,=\,\big(\mathscr{L}_m\, g_{\mu\nu} \!-\! T_{\mu\nu}\big) \cdot \bigg(f_{\!\mathscr{L}_m \mathscr{L}_m}\!\cdot \nabla^\mu \mathscr{L}_m + \sum_i  f_{\!\mathscr{L}_m \mathcal{R}_i}\!\cdot \nabla^\mu \mathcal{R}_i\bigg)\,,
\end{equation}
where $f_{\!\mathscr{L}_m \mathscr{L}_m}=\partial f_{\!\mathscr{L}_m}/\partial \mathscr{L}_m$ ,
$f_{\!\mathscr{L}_m \mathcal{R}_i}=\partial  f_{\!\mathscr{L}_m}/\partial \mathcal{R}_i$.
{Note that, if we adopt Eq.(\ref{AA second minimal coupling condition}) rather than Eq.(\ref{AA conjecture divergence eqn}) as the final form of nonconservation equation, the coefficient $(\mathscr{L}_m\, g_{\mu\nu} - T_{\mu\nu})=2\delta\mathscr{L}_m/\delta g^{\mu\nu}$ associated to the divergences
$\{\nabla^\mu \mathscr{L}_m\,, \nabla^\mu \mathcal{R}_i\}$
helps to clarify that they exclusively come from the $\mathscr{L}_m$-dependence in $\mathscr{L}=f$.}% \\

Following the weak conjecture, we now formally rewrite the divergence equation (\ref{Divergence of StressEnergyMomentum}) for $f(R,R_c^2,R_m^2,\mathscr{L}_m)$ gravity into
\begin{equation}\label{Divergence of StressEnergyMomentum Enu}
f_{\!\mathscr{L}_m}\! \cdot\nabla^\mu T_{\mu\nu}\,=\,\big(\mathscr{L}_m\, g_{\mu\nu} - T_{\mu\nu}\big)\,\nabla^\mu f_{\!\mathscr{L}_m}+\mathcal {E}_\nu \;,
\end{equation}
where
\begin{equation}\label{Enu to vanish}
\mathcal {E}_\nu\,\coloneqq\,-  f_{R_c^2} \nabla_\nu R_c^2- f_{R_m^2} \nabla_\nu R_m^2  + 2 \nabla^\mu  H_{\mu\nu}^{(\!f R_c^2)}+ 2 \nabla^\mu  H_{\mu\nu}^{(\!f R_m^2)}  \;,
\end{equation}
and $\mathcal {E}_\nu$ is expected to vanish by the weak conjecture, while $\mathcal {E}_\nu\equiv 0$ trivially holds under minimal coupling because of generalized Bianchi identities. Since we have not yet proved that $\mathcal {E}_\nu=0$, we preserve $\mathcal {E}_\nu$ in the divergence equation (\ref{Divergence of StressEnergyMomentum Enu}) and proceed to use it to check the equations of continuity and motion with different matter sources.

%%%%%%%%%%%%%%%%%%%%%%%%%%%%%%%%%%%%%%%%%%%%%%%%%%%%%%%%%%%%%%%%%%%%%%%%%%%%%
%%%%%%%%%%%%%%%%%%%%%%%%%%%%%%%%%%%%%%%%%%%%%%%%%%%%%%%%%%%%%%%%%%%%%%%%%%%%%

\section{Equations of continuity and nongeodesic motion}\label{Section 5}

Once the matter content in the spacetime is known, Eq.(\ref{Divergence of StressEnergyMomentum Enu}) can be concretized in accordance with the particular
forms of $T_{\mu\nu}$, which would imply the equations of continuity of the energy-matter content and the equation of (nongeodesic) motion for a test particle\footnote{The method and discussion in this section are also valid for a generic $\mathscr{L}\!=\!f(\mathcal{R}_1,\mathcal{R}_2\ldots,\mathcal{R}_n,\mathscr{L}_m)$ gravity as in Eq.(\ref{AA Tian-Booth weak conjecture condition}), and we just need to define the effective 1-form $\tilde{\mathcal{E}}_\nu=-\sum_i f_{\mathcal{R}_i}(\mathcal{R}_1\ldots\mathscr{L}_m)\cdot\!  \nabla_\nu \mathcal{R}_i + 2\sum_i  \nabla^\mu  H_{\mu\nu}^{(\!f R_i)}$ in place of the $\mathcal{E}_\nu$ for $f(R,R_c^2,R_m^2,\mathscr{L}_m)$ gravity. Specifically, $\tilde{\mathcal{E}}_\nu\equiv0$ under minimal coupling, and furthermore $\tilde{\mathcal{E}}_\nu$ vanishes universality if the weak conjecture were correct.}. This topic will be studied in this section, and note that  $T_{\mu\nu}$ and $\mathscr{L}_m$ will be adapted to the $(-,+++)$ metric signature.

\subsection{Perfect fluid}\label{subsection of Perfect fluid}
The stress-energy-momentum (SEM) tensor of a perfect fluid (no internal viscosity, no shear stresses, and zero thermal-conductivity coefficients)
with mass-energy density $\rho=\rho(x^\alpha)$, isotropic pressure $P=P(x^\alpha)$ and equation of state $P=w\,\rho$,
is given by\cite{Hawking Ellis}
\begin{equation}\label{SEM Tensor Perfect Fluid}
\begin{split}
T_{\mu\nu}^{\text{(PF)}} &=(\rho+P)\,u_\mu u_\nu +P\,g_{\mu\nu}\\
           &=\rho\,u_\mu u_\nu +P\,(g_{\mu\nu}+u_\mu u_\nu)\\
           &=\rho\,u_\mu u_\nu +P\,h_{\mu\nu}\;,\\
\end{split}
\end{equation}
where $u^\mu$ is the four-velocity along the worldline, satisfying $u_\mu u^\mu=-1$ and $u_\mu\!\nabla_{\!\nu} u^\mu=0$ ;
$h_{\mu\nu}$ is the projected spatial 3-metric, $h_{\mu\nu} \coloneqq g_{\mu\nu}+u_\mu u_\nu$ with inverse $h^{\mu\nu}=g^{\mu\nu}+u^\mu u^\nu$ , $h^{\mu\nu} u_\mu=0$, and $h^{\mu\nu}h_{\mu\nu}=3$.
Substituting Eq.(\ref{SEM Tensor Perfect Fluid}) into Eq.(\ref{Divergence of StressEnergyMomentum Enu}) and multiplying both sides by $u^\nu$, we get
\begin{equation}\label{Continuity PerfectFluid 1}
 u^\mu\,\nabla_\mu \rho+ (\rho+P)\, \nabla^\mu u_\mu \,=\,
 -(\mathscr{L}_m + \rho)\,u^\mu \,\nabla_\mu \ln f_{\!\mathscr{L}_m}
-f_{\!\mathscr{L}_m}^{-1}\,u^\nu \mathcal{E}_\nu\;,
\end{equation}
which generalizes the original continuity equation of perfect fluid in GR, $ u^\mu\nabla_\mu \rho+ (\rho+P) \nabla^\mu u_\mu =0$. % \\

On the other hand, after putting Eq.(\ref{SEM Tensor Perfect Fluid}) back to Eq.(\ref{Divergence of StressEnergyMomentum Enu}), use $h^{\xi\nu}$ to project the free index $\nu$, and it follows that
\begin{equation}
%\begin{split}
(\rho+P) \!\cdot\! u^\mu \nabla_\mu  u^\xi=
-h^{\xi\mu}\!\cdot\! \nabla_\mu P+h^{\xi\mu}\!\cdot\!   (\mathscr{L}_m-P) \,\nabla_\mu \ln f_{\!\mathscr{L}_m}
+ f_{\!\mathscr{L}_m}^{-1}\,  h^{\xi\nu}\mathcal{E}_\nu  \;,
%\end{split}
\end{equation}
where we have employed the properties $h^{\xi\nu}\!\cdot  u_\mu \nabla^\mu  u_\nu=g^{\xi\nu}\!\cdot  u_\mu \nabla^\mu  u_\nu=u_\mu \nabla^\mu  u^\xi$.
In general, $\rho+P\neq 0$ (in fact $\rho+P\geq 0$ by all four energy conditions in GR, and equality happens only for matters with large negative pressure). Thus we obtain the following absolute derivative along $u^\xi$ as the equation of motion:
\begin{equation}\label{EoM Perfect Fluid}
\frac{Du^\xi}{D\tau}\;\,\equiv\,\frac{du^\xi}{d\tau}+\Gamma^\xi_{\alpha\beta}u^\alpha u^\beta\,=\,a^{\,\xi}_{\text{(PF)}} + a^{\,\xi}_{(f_{\!\mathscr{L}_m})} +
a^{\,\xi}_{(\mathcal E)}\;,
\end{equation}
where $\tau$ is an affine parameter (e.g. proper time) for the timelike worldline along which $dx^\alpha=u^\alpha d\tau$, and the three proper accelerations are given by
\begin{equation} \label{Three Accelerations}
\hspace{1cm}\left\{ \begin{aligned}
a^{\,\xi}_{\text{(PF)}}\;\;&\equiv -h^{\xi\mu}\!\cdot   (\rho+P)^{-1}\,\nabla_\mu P \\
a^{\,\xi}_{(f_{\!\mathscr{L}_m}\!)}\,&\equiv- h^{\xi\mu}\!\cdot  (\rho+P)^{-1}\,\big(P-\mathscr{L}_m\big) \,\nabla_\mu \ln f_{\!\mathscr{L}_m} \\
a^{\,\xi}_{(\mathcal E)}\:\;\;\,&\equiv -h^{\xi\nu}\!\cdot  (\rho+P)^{-1} \,f_{\!\mathscr{L}_m}^{-1}\,\mathcal{E}_\nu\;.
\end{aligned} \right.
\end{equation}
Thus,  three proper accelerations are responsible for the nongeodesic motion. $a^{\,\xi}_{\text{(PF)}}$ is the standard acceleration from the pressure of fluid as in GR\cite{Hawking Ellis}, $a^{\,\xi}_{(f_{\!\mathscr{L}_m})}$ comes from the curvature-matter coupling,
while $a^{\,\xi}_{(\mathcal E)}$ is a collaborative effect of the \{$R_c^2$-, $R_m^2$-\}dependence in the action and their generic nonminimal coupling to $\mathscr{L}_m$. This is consistent with the result in \cite{f(RLm) Extra Force} in the absence of \{$R_c^2$, $R_m^2$\}. Also, all three accelerations are orthogonal to the worldline with tangent $u^\xi$, since
\begin{equation}
a^{\,\xi}_{\text{(PF)}}u_\xi\,=\,0\quad,\quad      a^{\,\xi}_{(f_{\!\mathscr{L}_m})} u_\xi\,=\,0\quad,\quad       a^{\,\xi}_{(\mathcal E)}u_\xi\,=\,0 \,.
\end{equation}

Both Eq.(\ref{Continuity PerfectFluid 1}) and Eqs.(\ref{EoM Perfect Fluid}) and (\ref{Three Accelerations}) depend on the choice of the perfect-fluid matter
Lagrangian density. If $\mathscr{L}_m=-\rho$ \cite{Hawking Ellis}\cite{Perfect fluid}, the continuity equation (\ref{Continuity PerfectFluid 1}) becomes
\begin{equation}\label{Continuity PerfectFluid 2}
u^\mu\,\nabla_\mu \rho+ (\rho+P)\, \nabla_\mu u^\mu \,=\,-f_{\!\mathscr{L}_m}^{-1}\,u^\nu \mathcal{E}_\nu\;,
\end{equation}
which is free from the gradient of the geometry-matter coupling strength $f_{\!\mathscr{L}_m}^{-1}\,u^\mu \nabla_\mu f_{\!\mathscr{L}_m}$, while
$a^{\,\xi}_{(f_{\!\mathscr{L}_m})}$ reduces to
\begin{equation}
 a^{\,\xi}_{(f_{\!\mathscr{L}_m})}\,\equiv- h^{\xi\mu}\!\cdot\! \,\nabla_\mu \ln f_{\!\mathscr{L}_m}\;,
\end{equation}
which does not rely on the equation of state $P=w\,\rho$.% \\

On the other hand, for the choice $\mathscr{L}_m=P$\cite{Perfect fluid}\cite{SotiriouFaraoniNongeodesic},
Eq.(\ref{Continuity PerfectFluid 1}) and Eq.(\ref{Three Accelerations}) respectively yields
\begin{equation}\label{Continuity PerfectFluid 3}
%\begin{split}
 u^\mu\,\nabla_\mu \rho+ (\rho+P)\, \nabla^\mu u_\mu \,=\,
 -( \rho+P)\,u^\mu \,\nabla_\mu \ln f_{\!\mathscr{L}_m}-f_{\!\mathscr{L}_m}^{-1}\,u^\mu \mathcal{E}_\mu\;,
%\end{split}
\end{equation}
and
\begin{equation}
 a^{\,\xi}_{(f_{\!\mathscr{L}_m})}\,\equiv\,0\;.
\end{equation}
Although the continuity equation (\ref{Continuity PerfectFluid 3}) looks pretty ordinary, the proper acceleration $a^{\,\xi}_{(f_{\!\mathscr{L}_m})}$ vanishes identically for $\mathscr{L}_m=P$ and consequently the nongeodesic motion in the gravitational field of the perfect fluid becomes independent of  the gradient of the nonminimal coupling strength $u^\mu \nabla_\mu f_{\!\mathscr{L}_m}$.% \\

%One can also takes into account the perfect-fluid equation of state $P=w\rho$, and use it wherever appropriate in these results.\\

As shown in \cite{Faraoni Nonminimal coupling II}, both $\mathscr{L}_m=P$ and $\mathscr{L}_m=-\rho$  are correct matter densities and both lead to the SEM tensor given in Eq.(\ref{SEM Tensor Perfect Fluid}). Differences of physical effects only occur in the situation of nonminimal coupling,
where $\mathscr{L}_m$ becomes a direct and explicit input in the energy-momentum divergence equation. In fact, as for the matter Lagrangian density $\mathscr{L}_m$ for a perfect fluid, one can also adopt the following ansatz,
\begin{equation}
\mathscr{L}_m\,=\,(a\rho+bP)\!\cdot\! g^{\alpha\beta} u_\alpha u_\beta + (c\rho+dP)\!\cdot\! g^{\alpha\beta}g_{\alpha\beta}\,=\,
(4c-a)\,\rho+(4d-b)\,P\;.
\end{equation}
Applying this to Eq.(\ref{Tmunu Equivalence}), the equality with Eq.(\ref{SEM Tensor Perfect Fluid}) yields  $a=-1/2=b$ and $c=-1/4=-d$, so
\begin{equation}
\mathscr{L}_m\,=\,\left(-\frac{1}{2}\rho-\frac{1}{2}P\right)\!\cdot\! g^{\alpha\beta} u_\alpha u_\beta +  \left(-\frac{1}{4}\rho+\frac{1}{4}P \right)\!\cdot\! g^{\alpha\beta}g_{\alpha\beta}\,=\,
-\frac{1}{2}\rho+\frac{3}{2}P\;.
\end{equation}
This density makes Eqs.(\ref{Continuity PerfectFluid 1}), (\ref{EoM Perfect Fluid}) and (\ref{Three Accelerations}) act normally, losing the aforementioned extraordinary properties associated with $\mathscr{L}_m=-\rho$ and $\mathscr{L}_m=P$.

%%%%%%%%%%%%%%%%%%%%%%%%%%%%%%%%%%%%%%%%%%%%%%%%%%%%%%%%%%%%%%%%%%%%%%%%%%%%%%%%%%%%%%%%%%
%%%%%%%%%%%%%%%%%%%%%%%%%%%%%%%%%%%%%%%%%%%%%%%%%%%%%%%%%%%%%%%%%%%%%%%%%%%%%%%%%%%%%%%%%%

\subsection{(Timelike) Dust}

The (timelike) dust source with mass-energy density  $\rho$ has SEM tensor\cite{Hawking Ellis}\cite{SotiriouFaraoniNongeodesic}
\begin{equation}\label{SEM Tensor Dust}
T_{\mu\nu}^{(\text{Dust})} \,=\,\rho\,u_\mu u_\nu\;,
\end{equation}
where $u_\mu=g_{\mu\nu}u^\nu$ with $u^\nu$ being the tangent vector field along the worldline of a timelike dust particle. One can still introduce the spatial metric $h_{\mu\nu}\equiv g_{\mu\nu}+u_\mu u_\nu$ orthogonal to $u^\mu$,
with $\{u_\mu\,,h_{\mu\nu}\}$ sharing all those properties as in the case of perfect fluid,
%With square $u_\nu u^\nu=-1$, thus $u_\mu\!\nabla_{\!\nu} u^\mu=0$. $h_{\alpha\nu}\,\equiv\,g_{\alpha\nu}+u_\alpha u_\nu$ and its inverse
%$h^{\alpha\nu}\,\equiv\,g^{\alpha\nu}+u^\alpha u^\nu$, $h^{\alpha\nu} u_\nu=0$, $h^{\alpha\nu}h_{\alpha\nu}=3$.
so dust acts  just like a perfect fluid with zero pressure, $P=0$. Substituting Eq.(\ref{SEM Tensor Dust}) back
into Eq.(\ref{Divergence of StressEnergyMomentum Enu}) and multiplying  by $u^\nu$ on both its sides yields
\begin{equation}
u^\mu \,\nabla_\mu \rho  +\rho\,\nabla^\mu  u_\mu \,=
-\big(\mathscr{L}_m+\rho\big)\, u^\nu \nabla_\nu \ln f_{\!\mathscr{L}_m} -f_{\!\mathscr{L}_m}^{-1}\,u^\nu \mathcal{E}_\nu\;,
\end{equation}
which modifies the continuity equation of dust $\nabla_\mu (\rho u^\mu) =0$ in GR. Meanwhile, projection of the free index $\nu$ by $h^{\xi\nu}$
in $\nabla^\mu T_{\mu\nu}^{(\text{Dust})}$  gives rise to the modified equation of  motion
\begin{equation}
\frac{Du^\xi}{D\tau}\;\,\equiv\,\frac{du^\xi}{d\tau}+\Gamma^\xi_{\alpha\beta}u^\alpha u^\beta\,=\,\hat{a}^{\,\xi}_{(f_{\!\mathscr{L}_m})} +
\hat{a}^{\,\xi}_{(\mathcal E)}\;,
\end{equation}
where
\begin{equation} \label{Two Accelerations Dust}
\hspace{0.5cm}\left\{ \begin{aligned}
\hat{a}^{\,\xi}_{(f_{\!\mathscr{L}_m})}\,&\equiv\,\;\; h^{\xi\mu}\!\cdot  \rho^{-1}\,\mathscr{L}_m \,\nabla_\mu \ln f_{\!\mathscr{L}_m} \\
\hat{a}^{\,\xi}_{(\mathcal E)}\;\;\;\,&\equiv -h^{\xi\nu}\!\cdot   \rho^{-1}\,f_{\!\mathscr{L}_m}^{-1}\,\mathcal{E}_\nu\;.
\end{aligned} \right.
\end{equation}
Being pressureless, the dust inherits just the two extra accelerations $\hat{a}^{\,\xi}_{(f_{\!\mathscr{L}_m})}$ and $\hat{a}^{\,\xi}_{(\mathcal E)}$, and both remain orthogonal to the worldline with tangent $u^\xi$,
\begin{equation}
\hat{a}^{\,\xi}_{(f_{\!\mathscr{L}_m})} u_\xi\,=\,0 \quad,\quad       \hat{a}^{\,\xi}_{(\mathcal E)}u_\xi\,=\,0\;.
\end{equation}

%%%%%%%%%%%%%%%%%%%%%%%%%%%%%%%%%%%%%%%%%%%%%%%%%%%%%%%%%%%%%%%%%%%%%%%%%
%%%%%%%%%%%%%%%%%%%%%%%%%%%%%%%%%%%%%%%%%%%%%%%%%%%%%%%%%%%%%%%%%%%%%%%%%

\subsection{Null dust}
The SEM tensor for null dust with energy density $\varrho$ is (e.g. \cite{SotiriouFaraoniNongeodesic})
\begin{equation}\label{SEM Tensor Null Dust}
T_{\mu\nu}^{\text{(ND)}}\,=\,\varrho\,\ell_\mu \ell_\nu\;,
\end{equation}
where $\ell_\mu=g_{\mu\nu}\ell^\nu$ with $\ell^\nu$ being the tangent vector field along the worldline of a null dust particle, $\ell_\mu\ell^\mu=0$.
$T_{\mu\nu}^{\text{(ND)}}$ together with the energy-momentum divergence equation (\ref{Divergence of StressEnergyMomentum Enu}) yields
\begin{equation}\label{ExpandedNullConservation}
 \ell_\nu\,\ell^\mu\nabla_\mu \varrho+  \varrho\, \ell^\mu\nabla_\mu  \ell_\nu +\varrho\,\ell_\nu \nabla_\mu   \ell^\mu \,=
\big(\mathscr{L}_m\, g_{\mu\nu} - \varrho\,\ell_\mu \ell_\nu \big)\,\nabla^\mu \ln f_{\!\mathscr{L}_m}+
f_{\!\mathscr{L}_m}^{-1}\,\mathcal{E}_\nu\;.
\end{equation}
Multiplying both sides with $\ell^\nu$, $\ell^\nu \ell_\nu=0$, $\ell_\nu\!\nabla_{\!\mu} \ell^\nu=0$, we obtain the following constraint:
\begin{equation}
f_{\!\mathscr{L}_m}\,\ell^\nu\nabla_\nu f_{\!\mathscr{L}_m}\,=\,
-\ell^\nu \mathcal{E}_\nu\;. \vspace{3mm}
\end{equation}

Now, introduce an auxiliary null vector field $n^\mu$ as null normal to $\ell^\mu$ such that $n^\mu n_\mu=0$, $\ell^\mu n_\mu=-1$, which induces the two-dimensional spatial metric $g_{\mu\nu}=-\ell_\mu n_\nu-n_\mu \ell_\nu +q_{\mu\nu}$, satisfying the conditions
\begin{equation}
q_{\mu\nu} q^{\mu\nu}=2 \quad, \quad q_{\mu\nu}\ell^\nu=0 = q_{\mu\nu} n^\nu \quad, \quad \ell^\alpha \nabla_\alpha q_{\mu\nu}=0\;.
\end{equation}
Multiplying Eq.(\ref{ExpandedNullConservation}) by $n^\nu$, and with $n^\nu\nabla_\mu \ell_\nu=-\ell^\nu\nabla_\mu n_\nu$,
we get the continuity equation
\begin{equation}\label{Continuity Null Dust}
\ell^\mu \nabla_\mu \varrho   +\varrho\, \nabla_\mu   \ell^\mu+  \varrho \ell^\nu \ell^\mu\nabla_\mu  n_\nu   \,=
-\big(\mathscr{L}_m\, n^\mu + \varrho\, \ell^\mu  \big)\,\nabla_\mu \ln f_{\!\mathscr{L}_m}-
f_{\!\mathscr{L}_m}^{-1}\,n^\nu \mathcal{E}_\nu\;,
\end{equation}
while projecting  Eq.(\ref{ExpandedNullConservation}) with $h^{\xi\nu}$ gives rise to the equation of motion along $\ell^\xi$,
\begin{equation}
%\begin{split}
\varrho \, \ell^\mu\nabla_\mu  \ell^\xi   \,=\,
\varrho \, \ell^\xi \ell^\nu \ell^\mu\nabla_\mu  n_\nu + h^{\xi\nu}\, \mathscr{L}_m\, \nabla_\nu \ln f_{\!\mathscr{L}_m}+
f_{\!\mathscr{L}_m}^{-1}\,h^{\xi\nu}\mathcal{E}_\nu\;,
%\end{split}
\end{equation}
\begin{equation}
\frac{D\ell^\xi}{D\lambda}\,\equiv\,\frac{d\ell^\xi}{d\lambda}+\Gamma^\xi_{\alpha\beta}\ell^\alpha \ell^\beta\,=\,\check{a}^\xi_{\text{(ND)}} + \check{a}^\xi_{(f_{\!\mathscr{L}_m})} +
\check{a}^\xi_{(\mathcal E)}\;,
\end{equation}
where $\lambda$ is an affine parameter for the null worldline along which $dx^\alpha=\ell^\alpha d\xi$,
and the three proper accelerations are respectively
\begin{equation} \label{Three Accelerations of Null Dust}
\left\{ \begin{aligned}
\check{a}^{\,\xi}_{\text{(ND)}}\;\;&\equiv\,\;\;  \ell^\xi \ell^\nu \ell^\mu\nabla_\mu  n_\nu \\
 \check{a}^{\,\xi}_{(f_{\!\mathscr{L}_m})}\,&\equiv\,\;\; h^{\xi\mu}\!\cdot     \varrho^{-1}\mathscr{L}_m\,\nabla_\nu \ln f_{\!\mathscr{L}_m} \\
\check{a}^{\,\xi}_{(\mathcal E)}\;\;\;\,&\equiv \,\;\;h^{\xi\nu}\!\cdot  \varrho^{-1}\,f_{\!\mathscr{L}_m}^{-1} \,\mathcal{E}_\nu\;.
\end{aligned} \right.
\end{equation}
As we can see, compared with timelike dust,   one more proper acceleration $\check{a}^{\,\xi}_{\text{(ND)}}$ shows up in the case of null dust,
and we will refer to it the \emph{affine} acceleration or \emph{inaffinity} acceleration.

%%%%%%%%%%%%%%%%%%%%%%%%%%%%%%%%%%%%%%%%%%%%%%%%%%%%%%%%%%%%%%%%%%%%%%%%%%%%%%%%%%%%%%%%%%
%%%%%%%%%%%%%%%%%%%%%%%%%%%%%%%%%%%%%%%%%%%%%%%%%%%%%%%%%%%%%%%%%%%%%%%%%%%%%%%%%%%%%%%%%%

\subsection{Scalar field}

The matter Lagrangian density and SEM tensor of a massive scalar field $\phi(x^\alpha)$ with mass $m$ in a potential $V(\phi)$ are respectively given by
\begin{equation}
\begin{split}
&\mathscr{L}_m\,=-\frac{1}{2}\big( \nabla_\alpha \phi \nabla^\alpha \phi + m^2 \phi^2\big)+V(\phi)\;,\\
T_{\mu\nu}\,=&\,\nabla_\mu \phi \nabla_\nu \phi-\frac{1}{2}g_{\mu\nu} \,\big( \nabla_\alpha \phi \nabla^\alpha \phi + m^2 \phi^2 -2V(\phi)\big)\;,
\end{split}
\end{equation}
thus $\mathscr{L}_m\, g_{\mu\nu} - T_{\mu\nu}=-\nabla_\mu \phi \nabla_\nu \phi$. For the $\nu$ component,
the equations of continuity and motion are both given by
\begin{equation}
\Big(\Box \phi  -m^2 \phi +V_\phi\Big)\cdot\!\nabla_\nu \phi\,=
 -\nabla_\nu \phi\cdot\!\nabla_\mu \phi \,\nabla^\mu \ln f_{\!\mathscr{L}_m}+ f_{\!\mathscr{L}_m}^{-1}\, \mathcal{E}_\nu \;.
\end{equation}
Specifically, by  setting $V(\phi)=0$ and under minimal coupling ($f_{\!\mathscr{L}_m}=$constant, $\mathcal{E}_\nu$=0), we get
\begin{equation}
\Box \phi  -m^2 \phi \,=\,0 \;,
\end{equation}
which is the standard covariant Klein-Gordon equation for spin-zero particles in  GR.\\

%%%%%%%%%%%%%%%%%%%%%%%%%%%%%%%%%%%%%%%%%%%%%%%%%%%%%%%%%%%%%%%%%%%%%%%%%%%%%%%%%%%%%%%%%%%%
%%%%%%%%%%%%%%%%%%%%%%%%%%%%%%%%%%%%%%%%%%%%%%%%%%%%%%%%%%%%%%%%%%%%%%%%%%%%%%%%%%%%%%%%%%%%
%%%%%%%%%%%%%%%%%%%%%%%%%%%%%%%%%%%%%%%%%%%%%%%%%%%%%%%%%%%%%%%%%%%%%%%%%%%%%%%%%%%%%%%%%%%%

\section{Further physical implications of nonminimal coupling}\label{Section 6}

We have seen that under nonminimal curvature-matter coupling, the divergence of the standard SEM density tensor is equal to the gradient of the coupling
strength $\nabla^\mu f_{\!\mathscr{L}_m}$ which, in general, will be nonvanishing. As such, the usual energy-momentum conservation laws for particular matter
fields will be modified as compared to the corresponding fields in general relativity. At the same time, as is discussed in  the Appendix , nonminimal coupling
also affects the energy conditions. The standard energy
energy conditions of general relativity are phrased in terms of the stress-energy tensor and require positive energies (null and strong) and causal flows of matter
(dominant). However, in applications these conditions are generally used to constrain the Riemann tensor and so the allowed geometries of spacetime
and structures like singularities or horizons. For standard general relativity the two approaches are essentially equivalent but for modified gravity they are not:
if the Einstein equations are modified then the
bounds on the Ricci tensor that achieve the desired effects generally do not translate into the usual restrictions on the stress-energy-momentum. Thus one is
faced with a choice: either keep the standard GR results and give up the usual energy conditions or keep the usual energy conditions but lose those results.

In this section we consider some immediate physical consequences of this choice. All of these are consequences of the Raychaudhuri equations for null
and timelike geodesic congruences and so the difference between the standard energy conditions and those needed to enforce the focussing theorems is
crucial to these discussions. These are considered in some detail in  the Appendix  and in the following $T_{\mu\nu}^{\text{(eff)}}$ refers to an effective
stress-energy tensor for which the standard form of the  energy conditions will leave those theorems intact.

\subsection{Black hole physics}

Many results in black hole physics follow from understanding a black hole horizon as a congruence of null geodesics whose evolution is governed by the
(twist-free) Raychaudhuri equation:
\begin{equation}
\frac{d\theta_{(\ell)} }{d\lambda}\,=\,\kappa_{(\ell)} \theta_{(\ell)} -\frac{1}{2}\,\theta^2_{(\ell)}-\sigma_{\mu\nu}^{(\ell)} \sigma^{\mu\nu}_{(\ell)}  -R_{\mu\nu}\ell^\mu \ell^\nu\;,
\label{ray}
\end{equation}
where $\ell^\mu = \left( \frac{\partial}{\partial \lambda} \right)^\mu$ is a null tangent to the horizon, and $\kappa_{(\ell)}$, $\theta_{(\ell)}$ and $\sigma^{(\ell)}_{\mu \nu}$
are respectively the associated acceleration/inaffinity, expansion and shear.

The second law of black hole mechanics follows from this equation along with the requirement that the congruence of null
curves that rules the event horizon have no future endpoints (see, for example, the discussion \cite{Hawking Ellis}). Now choosing an affine parameterization
for the congruence $\kappa_{(\ell)} = 0$ it is straightforward to see that the righthand side of (\ref{ray}) is nonpositive as long as $R_{\mu\nu}\ell^\mu \ell^\nu \geq0$.
In standard GR this follow from the null energy condition: $T_{\mu \nu} \ell^\mu \ell^\nu\geq 0$. It then almost immediately follows that $\theta_{(\ell)}$ must be
everywhere nonnegative.  Else  $\theta_{(\ell)} \rightarrow -\infty$ and the congruence focuses.
However, for modified gravity we will usually lose the equivalence $T_{\mu \nu} \ell^\mu \ell^\nu\geq 0 \Leftrightarrow R_{\mu\nu}\ell^\mu \ell^\nu \geq 0$
and so  we will be faced with a modified  area increase theorem if we require the standard energy conditions.

By similar arguments, again involving the null Raychaudhuri equation, the energy conditions play a crucial role in the theorems that require
trapped surfaces to be contained in black holes and singularities to lie in their causal future\cite{Hawking Ellis}. Thus for black hole physics, modifications of the energy conditions
are a serious business which can affect core results and intuitions.

\subsection{Wormholes}\label{Section Nonminimal Wormholes}

On the other hand, for those interested in faster-than-light travel changing the energy conditions would be a boon.  Introducing
the nonminimal gravitational coupling strength $f_{\!\mathscr{L}_m}$  brings new flexibility and the possibility of supporting wormholes, as shown in \cite{Generalized wormhole conditions} and \cite{Generalized wormhole conditions II} for a $\lambda R\!\cdot\mathscr{L}_m$ coupling term. More generally for the $\mathscr{L}\!=\!f(R,\mathcal{R}_1,\ldots,\mathcal{R}_n,\mathscr{L}_m)$ gravity, based on the generalized null and weak energy conditions developed in  the Appendix, it proves possible to defocus null and timelike congruences and form wormholes by violating these generalized conditions, while having the standard energy conditions in GR\cite{Hawking Ellis} maintained to exclude the need for exotic matters.
It also leads to an extra constraint $ f_{\!\mathscr{L}_m}/ f_R\geq 0$ as in Eq.(\ref{Generalized energy conditions Extra constraint}).

From Eq.(\ref{Generalized Null Energy Condition II}) in the Appendix, for a null congruence $\ell^\mu$,
one can maintain the standard null energy condition $T_{\mu\nu}\ell^\mu \ell^\nu\geq 0$ while violating
$T_{\mu\nu}^{\text{(eff)}}\ell^\mu \ell^\nu \leq 0$ (and so evade the focusing theorems) if
\begin{equation}\label{Generalized Null Defocusing Condition}
0\,\leq\,T_{\mu\nu}\,\ell^\mu \ell^\nu \,\leq\,2\,f_{\!\mathscr{L}_m}^{-1}\,\bigg(
\sum_i H_{\mu\nu}^{(\!f\mathcal{R}i)}\ell^\mu \ell^\nu - \ell^\nu \ell^\mu \nabla_\mu\!\nabla_\nu \,f_R \bigg) \;.
\end{equation}
Similarly for a timelike congruence, one has $T_{\mu\nu}\,u^\mu u^\nu\geq 0$ while $T_{\mu\nu}^{\text{(eff)}}\,u^\mu u^\nu \leq 0$, and Eq.(\ref{Generalized Weak Energy Condition}) leads to
\begin{equation}\label{Generalized Weak Defocusing Condition}
0\,\leq\,T_{\mu\nu}u^\mu u^\nu \,\leq\,f_{\!\mathscr{L}_m}^{-1}\,\bigg(f-R\,f_R
+2\sum_i H_{\mu\nu}^{(\!f\mathcal{R}i)} u^\mu u^\nu -2\,\big(u^\mu u^\nu \nabla_\mu\!\nabla_\nu+\Box\big)\,f_R \bigg)- \mathscr{L}_m\;.
\end{equation}
Specifically for $\mathscr{L}\!=\!f(R,R_c^2,R_m^2,\mathscr{L}_m)$ gravity, these two conditions are concretized as
\begin{equation}
0\,\leq\,T_{\mu\nu}\,\ell^\mu \ell^\nu \,\leq\,2\,f_{\!\mathscr{L}_m}^{-1}\,\bigg(
H_{\mu\nu}^{(\!fR_c^2)}\ell^\mu \ell^\nu+H_{\mu\nu}^{(\!fR_m^2)} \ell^\mu \ell^\nu - \ell^\nu \ell^\mu \nabla_\mu\!\nabla_\nu \,f_R \bigg) \; \;  \mbox{ and}
\end{equation}
\begin{equation}
0\,\leq\,T_{\mu\nu}u^\mu u^\nu \,\leq\,f_{\!\mathscr{L}_m}^{-1}\,\bigg(f-R\,f_R
+2H_{\mu\nu}^{(\!fR_c^2)} u^\mu u^\nu +2H_{\mu\nu}^{(\!fR_m^2)}u^\mu u^\nu  -2\,\big(u^\mu u^\nu \nabla_\mu\!\nabla_\nu+\Box\big)\,f_R \bigg)- \mathscr{L}_m\;,
\end{equation}
where $\{H_{\mu\nu}^{(\!fR_c^2)},\,H_{\mu\nu}^{(\!fR_m^2)}\}$ have been given in Eqs.(\ref{Vary Rc2}) and (\ref{Vary Rm2}).

Moreover, Eqs.(\ref{Generalized Null Defocusing Condition})(\ref{Generalized Weak Defocusing Condition}) indicate that in the case without  dependence on Riemannian invariants beyond $R$, i.e.
$\mathscr{L}=f(R,\mathscr{L}_m)$, a wormhole can be solely supported by the nonminimal-coupling effect if
\begin{equation}\label{wormhole by nonminimal coupling condition 1}
0\,\leq\,T_{\mu\nu}\,\ell^\mu \ell^\nu \,\leq\,-2\,f_{\!\mathscr{L}_m}^{-1}\,
  \ell^\nu \ell^\mu \nabla_\mu\!\nabla_\nu \,f_R  \; \;  \mbox{ and}
\end{equation}
%\begin{equation}
%\begin{split}&0\,\leq\,T_{\mu\nu}u^\mu u^\nu \,\leq\,- \mathscr{L}_m\,+\\
%f_{\!\mathscr{L}_m}^{-1}\,\Bigg(f-&Rf_R-2\big(u^\mu u^\nu \nabla_\mu\!\nabla_\nu+\Box\big)f_R \Bigg).
%\end{split} \end{equation}
\begin{equation}
%\begin{split}
0\,\leq\,T_{\mu\nu}u^\mu u^\nu \,\leq\,- \mathscr{L}_m\,+
f_{\!\mathscr{L}_m}^{-1}\,\Bigg(f-Rf_R
 -2\big(u^\mu u^\nu \nabla_\mu\!\nabla_\nu+\Box\big)f_R \Bigg).
%\end{split}
\end{equation}
For example, let $\mathscr{L}=f(R,\mathscr{L}_m)=R+2\kappa \mathscr{L}_m+\lambda\,R\mathscr{L}_m$, and the field equation (\ref{Gauss-Bonnet Best Simplified final form}) becomes
\begin{equation}
%\begin{split}
R_{\mu\nu}-\frac{1}{2}Rg_{\mu\nu}+\lambda\cdot\bigg(\mathscr{L}_m R_{\mu\nu}
+(g_{\mu\nu}\Box-\nabla_\mu\nabla_\nu)\mathscr{L}_m  \bigg)
=(\kappa+\frac{1}{2}\lambda R)T_{\mu\nu}
%\end{split}
\end{equation}
To have a quick realization of Eq.(\ref{wormhole by nonminimal coupling condition 1}), we further assume $\lambda=1$, $T_{\mu\nu}=\text{diag}[-\rho(r),P(r),P(r),P(r)]$, $\mathscr{L}_m=P(r)$ (recall Sec.~\ref{subsection of Perfect fluid}), and adopt the following simplest wormhole metric,
\begin{equation}
%\begin{split}
ds^2= -dt^2+dr^2+(r^2+L^2)\cdot\Big(d\theta^2 +\sin^2 \!\theta \,d\phi^2 \Big)\;,
%\end{split}
\end{equation}
with minimum throat scale $L$ and outgoing radial null vector field $\ell^\mu\partial_\mu=(-1,1,0,0)$. Then the condition Eq.(\ref{wormhole by nonminimal coupling condition 1}) reduces to become
\begin{equation}
%\begin{split}
0\,\leq\, -\rho+3P\,\leq\, \bigg(1+\frac{r^2}{L^2} \bigg)\,\partial_{r}\partial_{r} P\;,
%\end{split}
\end{equation}
which clearly shows that the standard null energy condition remains valid while spatial inhomogeneity of the pressure $\partial_{r}\partial_{r} P$ supports the wormhole.

Finally, note that it remains to be carefully checked whether solutions
exist that meet these conditions.

%%%%%%%%%%%%%%%%%%%%%%%%%%%%%%%%%%%%%%%%%%%%%%%%%%%%%%%%%%%%%%%%%%%%%%%%%%%%%%%%%%%%%%%%%%%%
%%%%%%%%%%%%%%%%%%%%%%%%%%%%%%%%%%%%%%%%%%%%%%%%%%%%%%%%%%%%%%%%%%%%%%%%%%%%%%%%%%%%%%%%%%%%

\section{Conclusions}

In this paper, we have derived the field equation for $\mathscr{L}\!=\!f(R,R_c^2,R_m^2,\mathscr{L}_m)$ fourth-order gravity
% with extra dependence on the Ricci square as opposed to GR, and allows
allowing for participation of the Ricci square $R_c^2$ and Riemann square $R_m^2$ in the Lagrangian density and nonminimal coupling between the curvature invariants
%$\{R\,,\,R_c^2\,,\,R_m^2\}$
and
%matter Lagrangian density
$\mathscr{L}_m$ as compared to GR. It turned out that $\mathscr{L}_m$ appears explicitly in the field equation because of confrontation between the nonminimal coupling and the traditional minimal definition of the SEM tensor $T_{\mu\nu}$.  When $f_{\mathscr{L}_m}=\text{constant}=2\kappa$, we recover the minimally coupled  $\mathscr{L}\!=\!f(R,R_c^2,R_m^2)+2\kappa\mathscr{L}_m$ model. Also, we have showed that both the curvature-$\mathscr{L}_m$ nonminimal coupling and the curvature-$T$ coupling are sensitive to the concrete forms of $\mathscr{L}_m$.

Secondly, by considering an explicit $R^2$-dependence, we have found the  smooth transition from $f(R,R_c^2,R_m^2,\mathscr{L}_m)$ gravity to the $\mathscr{L}\!=\!f(R,\mathcal{G},\mathscr{L}_m)$ generalized Gauss-Bonnet gravity by imposing the coherence condition $f_{R^2}\!=\!f_{R_m^2}\!=\! -f_{R_c^2}/4$. When $f(R,\mathcal{G},\mathscr{L}_m)$ reduces to the case $f(R,\mathscr{L}_m)+\lambda \mathcal{G}$ where $\mathcal{G}$ appears as a pure Gauss-Bonnet term, an extra term $\lambda\,\big(\!-\frac{1}{2}\mathcal{G}\,g_{\mu\nu}+2 R\,R_{\mu\nu}-4 R_\mu^{\;\;\,\alpha}R_{\alpha\nu}-4 R_{\alpha\mu\beta\nu}R^{\alpha\beta}
+2R_{\mu\alpha\beta\gamma}R_{\nu}^{\;\;\,\alpha\beta\gamma} \big)$ is left behind in the field equation representing the contribution from the covariant density $\lambda\!\sqrt{-g}\mathcal{G}$. We have shown that this term actually vanishes and thus $\lambda \mathcal{G}$ makes no difference to the gravitational field equation.

After studying the Gauss-Bonnet limit of  $f(R,R_c^2,R_m^2,\mathscr{L}_m)$ gravity, we moved on to more generic theories focusing on how the
%trying to understand  the fact
the standard stress-energy-momentum conservation equation $\nabla^\mu T_{\mu\nu}=0$ in GR is violated.  Under minimal coupling with $\mathscr{L}= f(\mathcal{R}_1,\ldots,\mathcal{R}_p)+2\kappa \mathscr{L}_m$, we commented that the generalized Bianchi identities and the Noether-induced definition of SEM tensor lead to automatic energy-momentum conservation. Under nonminimal coupling with $\mathscr{L}=f(\mathcal{R}_1,\ldots,\mathcal{R}_p,\mathscr{L}_m)$, we have proposed a weak conjecture and a strong one which state that the gradient of the nonminimal gravitational coupling strength $\nabla^\mu f_{\!\mathscr{L}_m}$ is the only divergence term balancing $f_{\mathscr{L}_m}\nabla^\mu T_{\mu\nu}$, while contributions from $\mathcal{R}_i$-dependence in the divergence equation all cancel out. Using the energy-momentum nonconservation equation specialized for  $f(R,R_c^2,R_m^2,\mathscr{L}_m)$ gravity, we have derived the equations of continuity and nongeodesic motion in the matter sources for perfect fluids, (timelike) dust, null dust, and massive scalar fields. % with s potential.
These equations directly generalize those in $f(\mathcal{R}_1,\ldots,\mathcal{R}_p,\mathscr{L}_m)$ gravity.

 Also, within $f(\mathcal{R}_1,\ldots,\mathcal{R}_p,\mathscr{L}_m)$ gravity, we have considered some  implications of nonminimal coupling and $\mathcal{R}_i$-dependence for black hole and wormhole physics. Moreover, it  is expected that the $\mathscr{L}=f(R,R_c^2,R_m^2, \mathscr{L}_m)$ model can provide many more possibilities to realize the late-time phase transition from cosmic deceleration to acceleration, and the energy-momentum nonconservation relation $f_{\!\mathscr{L}_m}\! \cdot\nabla^\mu T_{\mu\nu}=\big(\mathscr{L}_m g_{\mu\nu} - T_{\mu\nu}\big) \nabla^\mu f_{\!\mathscr{L}_m}$  under nonminimal coupling can cause interesting consequences in early-era cosmic evolution and compact astrophysical objects if is effective as a high-energy phenomenon. These topics will be extensively investigated in prospective studies.

\section*{Acknowledgement}

\noindent This work was financially supported by the Natural Sciences and Engineering Research
Council of Canada.

%%%%%%%%%%%%%%%%%%%%%%%%%%%%%%%%%%%%%%%%%%%%%%%%%%%%%%%%%%%%%%%%%%%%%%%%%%%%%%%%%%%%%%%%%%%%
%%%%%%%%%%%%%%%%%%%%%%%%%%%%%%%%%%%%%%%%%%%%%%%%%%%%%%%%%%%%%%%%%%%%%%%%%%%%%%%%%%%%%%%%%%%%

\appendix

\section{Generalized energy conditions for $f(R,\mathcal{R}_1,\ldots,\mathcal{R}_n,\mathscr{L}_m)$ gravity}
\label{AppA}

For the generic $\mathscr{L}=f(R,\mathcal{R}_1,\ldots,\mathcal{R}_n,\mathscr{L}_m)$ gravity introduced in Section 4,
the variational principle or equivalently  $\displaystyle \frac{1}{\sqrt{-g}}\frac{\delta\, (\!\sqrt{-g}\,\mathscr{L})}{\delta g^{\mu\nu}}=0$ yields the field equation:
\begin{equation}\label{Appendix Generic Field Eqn}
-\frac{1}{2}fg_{\mu\nu}+f_R\,R_{\mu\nu}+ \big(g_{\mu\nu}\Box - \nabla_\mu\!\nabla_\nu\big)\,f_R+\sum_i H_{\mu\nu}^{(\!f{\mathcal{R}}_i)}  \,=\,\frac{1}{2}f_{\!\mathscr{L}_m}\!\cdot  \big(T_{\mu\nu}-\mathscr{L}_m\, g_{\mu\nu}\big)\;,
\end{equation}
where $H_{\mu\nu}^{(\!f{\mathcal{R}}_i)}\!\cdot\delta g^{\mu\nu}\!\coloneqq\!f_{\mathcal{R}_i}\!\cdot\delta \mathcal{R}_i$. An immediate and very useful implication of this field equation is a group of generalized null, weak, strong and dominant energy conditions (abbreviated into NEC, WEC, SEC and DEC respectively), which has been employed in Sec.~\ref{Section Nonminimal Wormholes} in studying effects of nonminimal coupling in supporting wormholes.

Recall that in a (region of) spacetime filled by a null or a timelike congruence, the expansion rate along the null tangent $\ell^\mu$ or
the timelike tangent $u^\mu$ is given by the respective Raychaudhuri equation\cite{Hawking Ellis}:
\begin{equation}
\ell^\mu \nabla_\mu \theta_{(\ell)} \,=\,\frac{d\theta_{(\ell)} }{d\lambda}\,=\,\kappa_{(\ell)} \theta_{(\ell)} -\frac{1}{2}\,\theta^2_{(\ell)}-\sigma_{\mu\nu}^{(\ell)} \sigma^{\mu\nu}_{(\ell)} +\omega_{\mu\nu}^{(\ell)} \omega^{\mu\nu}_{(\ell)} -R_{\mu\nu}\ell^\mu \ell^\nu\; \; \mbox{and}
\end{equation}
\begin{equation}
\hspace{-4.6mm}u^\mu \nabla_\mu \theta_{(u)}\,=\,\frac{d\theta_{(u)}}{d\tau}\,=\,\kappa_{(u)}  \theta_{(u)}-\frac{1}{3}\,\theta^2_{(u)}-\sigma_{\mu\nu}^{(u)}\sigma^{\mu\nu}_{(u)}+\omega_{\mu\nu}^{(u)}\omega^{\mu\nu}_{(u)}-R_{\mu\nu}u^\mu u^\nu\;.
\end{equation}
Under affine parametrizations one has $\kappa_{(\ell)}=0=\kappa_{(u)}$,  for hypersurface-orthogonal congruences the twist vanishes $\omega_{\mu\nu}\omega^{\mu\nu}=0$, and the shear as a spatial tensor ($\sigma_{\mu\nu}^{(\ell)}\ell^\mu=0$, $\sigma_{\mu\nu}^{(u)}u^\mu=0$) always satisfies $\sigma_{\mu\nu}\sigma^{\mu\nu}\geq 0$.
Thus, to ensure $d\theta_{(\ell)}/d\lambda\leq 0$ and $d\theta_{(u)}/d\tau\leq 0$ under all conditions so that ``gravity always gravitates'' and the congruence focuses, the following geometric nonnegativity conditions should hold:
\begin{align}\label{Generalized EC Geometri conditions}
R_{\mu\nu}\ell^\mu \ell^\nu \,\geq\,0\quad (\text{NEC})\qquad,\qquad
R_{\mu\nu}u^\mu u^\nu \,\geq\,0\quad (\text{SEC})\;.
\end{align}
On the other hand, the field equation (\ref{FieldEq-2}) can be recast into a compact GR form,
\begin{equation}\label{FieldEqnGRForm}
G_{\mu\nu} \equiv R_{\mu\nu}-\frac{1}{2}Rg_{\mu\nu} = \kappa \,T_{\mu\nu}^{\text{(eff)}}\;,\qquad
R =-\kappa\, T^{\text{(eff)}}\;,\qquad   R_{\mu\nu} = \kappa\, \Big(T_{\mu\nu}^{\text{(eff)}}-\frac{1}{2} g_{\mu\nu}T^{\text{(eff)}} \Big)\;,
\end{equation}
where all terms beyond  GR ($G_{\mu\nu}=\kappa T_{\mu\nu}$) in Eq.(\ref{Appendix Generic Field Eqn}) have been  packed into the effective SEM tensor $T_{\mu\nu}^{\text{(eff)}}$,
\begin{equation}\label{Our Effective Tmunu General}
T_{\mu\nu}^{\text{(eff)}}\,=\,\frac{1}{2\kappa }\,\frac{f_{\!\mathscr{L}_m}}{f_R}\, \bigg(T_{\mu\nu}-\mathscr{L}_m\, g_{\mu\nu}\bigg)+ \frac{1}{2\kappa }\,\frac{f_{\!\mathscr{L}_m}}{f_R}\,\Bigg((f-R\,f_R)\,g_{\mu\nu} +2\big(\nabla_\mu\!\nabla_\nu-g_{\mu\nu}\Box\big)\,f_R -2\sum_i H_{\mu\nu}^{(\!f\mathcal{R}i)}  \Bigg)\;.
\end{equation}
The purely \emph{geometric}  conditions  Eq.(\ref{Generalized EC Geometri conditions}) can be translated into \emph{matter} nonnegativity  conditions through Eq.(\ref{FieldEqnGRForm}),
\begin{align}\label{Generalized energy conditions}
T_{\mu\nu}^{\text{(eff)}}\,\ell^\mu \ell^\nu \,\geq\,0\quad (\text{NEC})\quad,\quad
T_{\mu\nu}^{\text{(eff)}}u^\mu u^\nu\,\geq\,\frac{1}{2}\, T^{\text{(eff)}}u_\mu u^\mu
\quad (\text{SEC})\quad,\quad
T_{\mu\nu}^{\text{(eff)}}\,u^\mu u^\nu \,\geq\,0\quad (\text{WEC})\;,
\end{align}
where $u_\mu u^\mu=-1$  in SEC for the signature $(-,+++)$ used in this paper.
Then the generalized NEC in Eq.(\ref{Generalized energy conditions}) is expanded into (as $\kappa>0$)
\begin{equation}\label{Generalized Null Energy Condition}
\frac{f_{\!\mathscr{L}_m}}{ f_R}\, T_{\mu\nu}\,\ell^\mu \ell^\nu +\frac{2}{f_R}\,\bigg(
\ell^\nu \ell^\mu \nabla_\mu\!\nabla_\nu \,f_R -\sum_i H_{\mu\nu}^{(\!f\mathcal{R}i)} \ell^\mu \ell^\nu \bigg)
\;\geq\,0\;,
\end{equation}
which is the simplest one with $\mathscr{L}_m$ absent. Now, consider a special situation where $f_R=$constant and $H_{\mu\nu}^{(\!f\mathcal{R}i)}\!=\!0$ (i.e. dropping all dependence on $\mathcal{R}_i$ in $f$), so Eq.(\ref{Generalized Null Energy Condition}) reduces to $\Big( f_{\!\mathscr{L}_m}/ f_R\Big)\cdot\! T_{\mu\nu}\ell^\mu \ell^\nu\geq 0$; since $T_{\mu\nu}\ell^\mu \ell^\nu\geq 0$ due to the standard NEC in GR, which continues to hold here as exotic matters are unfavored, we obtain an extra constraint
\begin{equation}\label{Generalized energy conditions Extra constraint}
\frac{f_{\!\mathscr{L}_m}}{ f_R}\,\geq\,0 \;,
\end{equation}
with which  Eq.(\ref{Generalized Null Energy Condition}) becomes
\begin{equation}\label{Generalized Null Energy Condition II}
T_{\mu\nu}\,\ell^\mu \ell^\nu +2\,f_{\!\mathscr{L}_m}^{-1}\,\bigg(
\ell^\nu \ell^\mu \nabla_\mu\!\nabla_\nu \,f_R- \sum_i H_{\mu\nu}^{(\!f\mathcal{R}i)} \ell^\mu \ell^\nu  \bigg)\,\geq\,0 \;,
\end{equation}
and the WEC in Eq.(\ref{Generalized energy conditions}) can be expanded into
\begin{equation}\label{Generalized Weak Energy Condition}
%\begin{split}
T_{\mu\nu}u^\mu u^\nu + \mathscr{L}_m+  f_{\!\mathscr{L}_m}^{-1}\,\bigg(R\,f_R-f
+2\big(u^\mu u^\nu \nabla_\mu\!\nabla_\nu+\Box\big)\,f_R -2\sum_i H_{\mu\nu}^{(\!f\mathcal{R}i)}u^\mu u^\nu  \bigg)\,\geq\,0\;.
%\end{split}
\end{equation}
In general, the pointwise nonminimal coupling strength $f_{\!\mathscr{L}_m}$ can take either positive or negative values. However, recall that within $f(R)+2\kappa\mathscr{L}_m$ gravity, physically viable models specializing $f(R)$ should satisfy
$f_R > 0$ and $f_{RR} > 0$ \cite{f(R) gravity Review by Sotiriou and Faraoni}; if this were still true in $f(R,\mathcal{R}_1,\ldots,\mathcal{R}_n,\mathscr{L}_m)$ gravity, we would get $f_{\!\mathscr{L}_m}>0$ by the extra constraint Eq.(\ref{Generalized energy conditions Extra constraint}), which would be in strong agreement with the case of minimal coupling when $f_{\!\mathscr{L}_m}=2\,\kappa>0$.

Applying Eqs.(\ref{Our Effective Tmunu General}), (\ref{Generalized Null Energy Condition II}) and (\ref{Generalized Weak Energy Condition}) to  the Lagrangian density $\mathscr{L}\!=\!f(R,R_c^2,R_m^2,\mathscr{L}_m)$, we immediately obtain
\begin{equation}\label{Our Effective Tmunu  Rc2 Rm2}
T_{\mu\nu}^{\text{(eff)}}\,=\,\frac{1}{2\kappa }\,\frac{f_{\!\mathscr{L}_m}}{f_R}\, \bigg(T_{\mu\nu}-\mathscr{L}_m\, g_{\mu\nu}\bigg)+ \frac{1}{2\kappa }\,\frac{f_{\!\mathscr{L}_m}}{f_R}\,\Bigg((f-R\,f_R)\,g_{\mu\nu} +2\big(\nabla_\mu\!\nabla_\nu-g_{\mu\nu}\Box\big)\,f_R -2H_{\mu\nu}^{(\!fR_c^2)} -2H_{\mu\nu}^{(\!fR_m^2)}  \Bigg)\;.
\end{equation}
as the effective SEM tensor for for $f(R,R_c^2,R_m^2,\mathscr{L}_m)$ gravity.
Then relative to the standard SEM tensor the generalized null and weak energy conditions respectively become
\begin{equation}\label{Generalized Null Energy Condition II Rc2 Rm2}
T_{\mu\nu}\,\ell^\mu \ell^\nu +2\,f_{\!\mathscr{L}_m}^{-1}\,\bigg(
\ell^\nu \ell^\mu \nabla_\mu\!\nabla_\nu \,f_R- H_{\mu\nu}^{(\!fR_c^2)} \ell^\mu \ell^\nu - H_{\mu\nu}^{(\!fR_m^2)} \ell^\mu \ell^\nu  \bigg)\,\geq\,0 \;\;\text{ and}
\end{equation}
\begin{equation}\label{Generalized Weak Energy Condition Rc2 Rm2}
%\begin{split}
T_{\mu\nu}u^\mu u^\nu + \mathscr{L}_m+  f_{\!\mathscr{L}_m}^{-1}\,\bigg(R\,f_R-f
+2\big(u^\mu u^\nu \nabla_\mu\!\nabla_\nu+\Box\big)\,f_R -2H_{\mu\nu}^{(\!fR_c^2)}u^\mu u^\nu-2H_{\mu\nu}^{(\!fR_m^2)}u^\mu u^\nu  \bigg)\,\geq\,0\;,
%\end{split}
\end{equation}
where $\{H_{\mu\nu}^{(\!fR_c^2)},\,H_{\mu\nu}^{(\!fR_m^2)}\}$ have been given in Eqs.(\ref{Vary Rc2}) and (\ref{Vary Rm2}).

Also, with Eq.(\ref{Our Effective Tmunu General}) one can directly obtain the concrete forms SEC and DEC for $\mathscr{L}\!=\!f(R,\mathcal{R}_1,\ldots,\mathcal{R}_n,\mathscr{L}_m)$ gravity,
which however will not be listed here.

%%%%%%%%%%%%%%%%%%%%%%%%%%%%%%%%%%%%%%%%%%%%%%%%%%%%%%%%%%%%%%%%%%%%%%%%%%%%%%%%%%%%%%%%%%%%
%%%%%%%%%%%%%%%%%%%%%%%%%%%%%%%%%%%%%%%%%%%%%%%%%%%%%%%%%%%%%%%%%%%%%%%%%%%%%%%%%%%%%%%%%%%%

\end{document}